\documentclass[pra,aps,a4paper,twocolumn,amsmath,amssymb,floatfix,nobalancelastpage,superscriptaddress,notitlepage,accepted=2023-04-06]{quantumarticle}
\pdfoutput=1
\usepackage{graphics,amssymb,amsmath,epsfig,color,textgreek}
\usepackage{graphicx}
\usepackage[dvipsnames]{xcolor}
\usepackage{dcolumn}
\usepackage{bm}
\usepackage[colorlinks=true,citecolor=cyan]{hyperref}
\hypersetup{colorlinks=true,citecolor=cyan,linkcolor=red,urlcolor=magenta}
\usepackage{braket}
\usepackage[normalem]{ulem}
\usepackage{cancel}
\usepackage{diagbox}
\usepackage{soul}

\newcommand{\beq}{\begin{equation}}
\newcommand{\eeq}{\end{equation}}

\newcommand{\bk}{{\bm k}}

\begin{document}

\preprint{}
\title{Robust measurement of wave function topology on NISQ quantum computers}

\author{Xiao Xiao}
\email{phxxiao@gmail.com}
\affiliation{Department of Physics, North Carolina State University, Raleigh, North Carolina 27695, USA}

\author{J.~K.~Freericks}
\email{james.freericks@georgetown.edu}
\affiliation{Department of Physics, Georgetown University, 37th and O Sts. NW, Washington, DC 20057 USA}

\author{A.~F.~Kemper}
\email{akemper@ncsu.edu}
\affiliation{Department of Physics, North Carolina State University, Raleigh, North Carolina 27695, USA}

\begin{abstract}
Topological quantum phases of quantum materials are defined through their topological invariants. These topological invariants are quantities that characterize the global geometrical properties of the quantum wave functions and thus are immune to local noise. Here, we present a strategy to measure topological invariants on quantum computers. We show that our strategy can be easily integrated with the variational quantum eigensolver (VQE) so that the topological properties of generic quantum many-body states can be characterized on current quantum hardware. We demonstrate the robust nature of the method by measuring topological invariants for both non-interacting and interacting models, and map out interacting quantum phase diagrams on quantum simulators and IBM quantum hardware.

\end{abstract}
\vspace{1.0cm}

\maketitle

\section{introduction}

Topological phases 
are characterized by nonlocal topological invariants, which are by nature robust against local perturbations \cite{Thouless_prl_1982,Niu_prb_1985,Sheng_PRB_2006,Obuse_PRB_2007,Li_PRL_2009,Prodan_JPA_2011,Chalker_PRB_2011,Liu_PRL_2012,Lobos_PRL_2012,Konig_PRB_2013,Altland_PRL_2014,Shem_PRL_2014,Song_PRB_2014,Foster_PRB_2014,Wang_PRB_2014,Liu_PRL_2017,Meier_Science_2018,Stutzer_Nature_2018,Xiao_arXiv_2018,Shtanko_PRL_2018,Okugawa_PRB_2020}. 
This unique property makes determining properties of topological phases an ideal application of quantum computing in the noise intermediate-scale quantum (NISQ) era, where the noise levels are high.
A significant amount of work has been performed on realizing topological phases and identifying different topological phases \emph{qualitatively} on quantum hardware \cite{Roushan_Nature_2014,Choo_PRL_2018,Smith_arXiv_2019,Azses_PRL_2020,Mei_PRL_2020,Xiao_arXiv_2020}. Nevertheless, although the strategies for calculating topological invariants are well-established in the condensed matter community,  there have only been a few studies employing quantum circuits to determine them \cite{Flurin_PRX_2017,Zhan_PRL_2017,Xu_PRL_2018,Elben_SA_2020}. 

\begin{figure}[t]
  \centering
  \includegraphics[width=0.9\columnwidth]{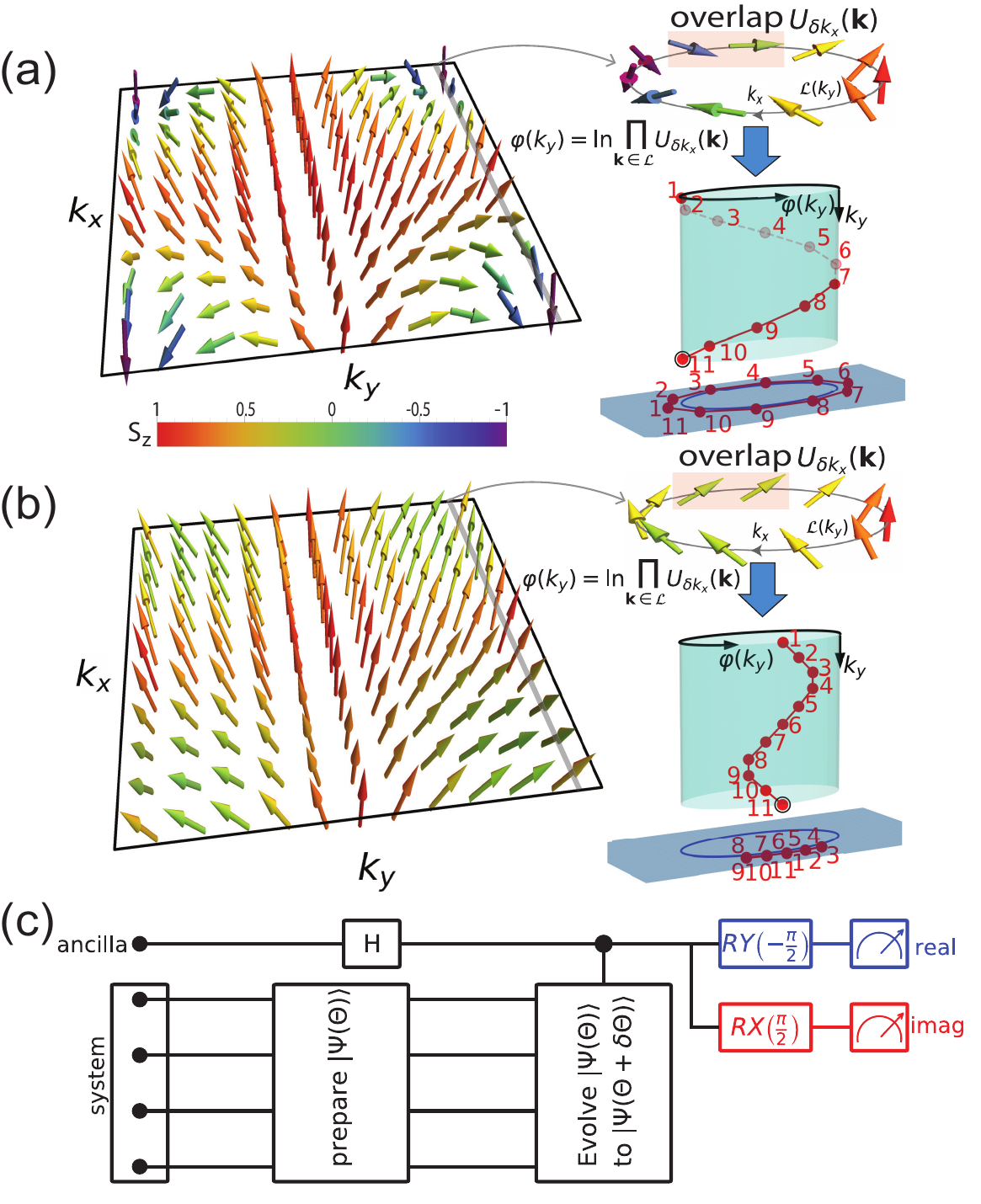}
  \caption{Illustration of different topology of wave functions: the pseudo-spin representation of the wave function of a chiral $p$-wave superconductor in (a) the case of a topological state and (b) the case of a trivial state. 
  The pseudo-spin vector field for a particular $k_y$ (along the gray lines) determines the Zak phase $\varphi(k_y)$, which has non-trivial (trivial) winding for the topological (trivial) state. The central quantity here is the overlap of the wave function after a small transport in $\bk$-space denoted by $U_{\delta \bk}(\bk)$, which can be measured by the generic quantum circuit shown in (c).}
   \label{fig1}
\end{figure}

The difficulty in using NISQ hardware to measure topological invariants stems from the inherent errors due to the non-fault-tolerant quantum hardware; the issues of noise are omnipresent within NISQ hardware calculations \cite{Preskill_Quantum_2018}, and advanced error mitigation strategies often have to be deployed before even \emph{semiquantitative} results are obtained \cite{Kandala_Nature_2019,Hamilton_QMI_2020}. These strategies may not suffice; the quantitative results may  differ significantly from the exact results, regardless of error mitigation. Even to obtain qualitatively correct results, limitation to a few qubits and low-depth circuits \cite{Lierta_Quantum_2018,Aydeniz_npj_2020} is usually necessary to reduce the influence of the gate errors in NISQ quantum computers.

Here we develop quantum circuits\textemdash based on holonomy\textemdash that can measure topological invariants of models, and do so in a error-resistant (or even error-free) manner. Our strategy is to construct a general quantum circuit to measure the parallel transport of wave functions in the base space. This determines the connection of the wave function bundle, which permits the gauge-invariant calculation of
topological invariants. Importantly, by relying on global properties of the wave function, an integer result can be obtained without any rounding as long as the errors acquired from the quantum hardware fall below a maximal noise threshold.

We first demonstrate our strategy by calculating different topological invariants for chiral $p$-wave superconductors \cite{Volovik_JEPT_1999,Read_PRB_2000}, which works with a single-particle wave function and an exact quantum circuit. Then, to further validate the general application of our strategy, we focus on the calculations of Chern numbers and calculated them for a quantum Hall state~\cite{Fukui_JPSJ_2005} where the wave functions are prepared variationally (via adaptive VQE). This demonstrates that our approach is broadly applicable with an affordable depth on NISQ hardware \cite{Malley_PRX_2016, Kandala_Nature_2017,Kokail_Nature_2019,Yuan_quantum_2019,Grimsley_NC_2019}. More importantly, our strategy in combination with VQE even allows us to correctly calculate Chern numbers for interacting models, and we support this point by calculating Chern numbers for the quantum Hall state \cite{Fukui_JPSJ_2005} with Hatsugai-Kohmoto type interactions \cite{Hatsugai_JPSJ_1992,Kudo_PRL_2019,Philips_NP_2020} and an interacting Chern insulator model \cite{Qi_PRB_2006,Wu_PRL_2016} on quantum hardware and quantum simulators respectively.
Strikingly, the Chern numbers for various models (even with interactions) can be calculated \emph{exactly} on NISQ machines without any error. We are not aware of any other error-free measurements obtained from NISQ hardware. Our results provide remarkable examples of the robustness of determining topological properties on NISQ machines.

The paper is organized as follows. Our discussion begins with the description of the general scheme to measure a topological invariant in quantum hardware in Sec.~\ref{sec2}. The general quantum circuit and method to realize the measurement of the topological invariants are provided and discussed in this section. Following this, we demonstrate the concrete applications of this general scheme with detailed examples. In Sec.~\ref{sec3} we consider a model of chiral $p$-wave topological superconductors, and we demonstrate how to construct the exact quantum circuit to measure the assocated topological invariants for this model on real quantum hardware, which includes the Chern number, the winding of Zak phases and the ensemble geometric phase \cite{Bardyn_PRX_2018}. Within this context, in Sec.~\ref{sec:calc_chern},
we also discuss how this formalism can be used to obtain a robust, integer-valued Chern number without any rounding.

The exact circuits discussed in Sec.~\ref{sec3} can be constructed only when the model can be solved exactly. To go beyond this limitation, we integrate our measurement scheme with VQE, which allows us to measure topological invariants for arbitrary quantum states including interacting models. To this end, we demonstrate this strategy with three examples: the calculation of Chern numbers for the flux-$2\pi/3$ quantum hall model in Sec.~\ref{sec4a}, the determination of the interacting topological phase diagrams for the flux-$2\pi/3$ quantum hall model with Hatsugai-Kohmoto type interactions in Sec.~\ref{sec4b}, and the determination of the interacting topological phase diagrams for an interacting Chern insulator model in Sec.~\ref{sec4c}. In Sec.~\ref{sec5} we contextualize our results by discussing the differences from other existing proposals and the potential for calculating topological invariants beyond the capability of classical computers with this method.

\section{General scheme and quantum circuit} \label{sec2}

To measure the wave function topology in some parameter space, the central quantity is the holonomy in the wave function bundle, obtained by the parallel transportation of the wave function along a closed loop in the base space. Topological states have non-trivial holonomy, while trivial states have trivial holonomy (see Fig.~\ref{fig1}). Parallel transport reveals a local connection, which is determined from the overlap of wave functions $\langle \Psi_{\Theta}|\Psi_{\Theta+\delta\Theta} \rangle$ at two neighboring points $\Theta$ and $\Theta+\delta\Theta$ in the base space. Therefore, the key step to measure the holonomy is to determine the local wave function overlap, which requires
the evolution of the wave function from $|\Psi_{\Theta}\rangle$ to $|\Psi_{\Theta+\delta\Theta} \rangle$. Once this is known, the overlap $\langle \Psi_{\Theta}|\Psi_{\Theta+\delta\Theta} \rangle$ can be evaluated by a Hadamard test (see Fig.~\ref{fig1} (c)) \cite{Aharonov_Algo_2009,Murta_PRA_2020}. 

To make the discussion in the above concrete, we consider models defined in a two-dimensional space. Because we are interested in bulk properties, periodic boundary conditions are applied. Then the topological information of the wave function is stored in the two-dimensional (magnetic) Brillouin zones (BZs). To implement the holonomy operation, the BZs are discretized into a $N_x\times N_y$ mesh with the mesh points in the BZs denoted as $\bk = (k_x,k_y)$. Then the local connection at a particular mesh point $\bk$ is characterized by the normalized overlap of the wave function at $\bk$ and its neighbors in the discretized BZ. We denote the normalized overlap at a particular point $\bk$ as $U_{\delta \bk}(\bk) \equiv \langle \Psi(\bk)|\Psi(\bk+\delta\bk) \rangle/|\langle \Psi(\bk)|\Psi(\bk+\delta\bk) \rangle|$, where $\delta \bk = \delta k_x \hat{x}$ or $\delta \bk = \delta k_y \hat{y}$ with $\hat{x}$ and $\hat{y}$ the unit vectors along the $x$- and $y$-directions. Once $U_{\delta \bk}(\bk)$ for all the mesh point in the BZ are measured, different kinds of topological invariants can be constructed from them. In this work, we will demonstrate the measurements of three topological invariants: the Chern number, the Zak phase, and the ensemble geometric phase. 

\subsection{Chern number}

First, in a discretized 2$D$ BZ the Chern number can be expressed as \cite{Fukui_JPSJ_2005}:
\begin{align} \label{Eq1}
\mathcal{C} = \frac{1}{2\pi i} \sum_{\bk} \mathcal{F}(\bk),
\end{align}
where the local gauge field $\mathcal{F}(\bk)$ (or Berry curvature) is defined from the normalized overlap $U_{\delta \bk}(\bk)$ within a plaquette formed by the neighboring four mesh points:
\begin{align}
\label{Eq2}
\mathcal{F}(\bk) = \ln \left[  \frac{U_{\delta \hat{\bk}_x} (\bk) U_{\delta \hat{\bk}_y}(\bk+\delta \hat{\bk}_x ) }{U_{\delta \hat{\bk}_x}(\bk+\delta \hat{\bk}_y ) U_{\delta \hat{\bk}_y}(\bk)} 
\right].
\end{align}
Here $\delta \hat{\bk}_x = \delta k_x \hat{x}$ ($\delta \hat{\bk}_y = \delta k_y \hat{y}$) denotes the grid spacing of the discretized BZ along the $x$ ($y$) direction. The above analysis indicates that $U_{\delta \bk}(\bk)$ is indeed central in the measurement of Chern numbers on quantum hardware.
In Sec.~\ref{sec:calc_chern}, we will outline
how this formalism can be used to obtain a robust, integer-valued Chern number without any rounding.

\subsection{Zak phase}

Another way to determine the topology of the wavefunction is to measure the winding of the Zak phase, which is defined by parallel transport of the wavefunction in only one direction of the Brillouin zone. The Zak phase measures the holomony of the wavefunction in one direction and can be written in terms of normalized overlap $U_{\delta \hat{\bk}_x}(\bk)$ as:
\beq \label{Zak}
\varphi(k_y) = \ln \prod_{\bk \in \mathcal{L}(k_y)} U_{\delta k_x}(\bk),
\eeq
where $\mathcal{L}(k_y)$ is the loop for a fixed $k_y$ along the $k_x$-direction. 
Then, as illustrated in Fig.~\ref{fig1}(a) and (b), the winding of the Zak phase along the $k_y$-direction is related to the Chern number as follows:
\beq
\mathcal{C} = \frac{1}{2\pi} \oint dk_y \frac{d \varphi(k_y)}{d k_y}.
\eeq

\subsection{Ensemble Geometric Phase}

From the normalized overlap $U_{\delta \bk}(\bk)$, we can even obtain the topology of a mixed state density matrix $\rho$; these may arise from a finite temperature or by being driven out of equilibrium. Here we focus on the ensemble geometric phase $\varphi_E$ \cite{Bardyn_PRX_2018} of a mixed state in non-interacting systems without particle number fluctuations. The ensemble geometric phase can be viewed as a many-body generalization of Zak phase and is defined as:
\beq
\varphi_E(k_y) = \Im m \left[\ln \langle e^{i \delta k_x \hat{X}} \rangle \right],
\eeq
where $\langle \cdots \rangle = \mathrm{Tr}[\hat{\rho}\cdots]$ with $\hat{\rho}$ the density matrix of the system, and $\hat{X} = \sum_{j} \hat{x}_j$ is the many-body position operator with $\hat{x}_j$ the position operator for the $j$th particle. In real space, the density matrix $\hat{\rho}$ can be expressed as:
\beq
\hat{\rho} = \frac{1}{Z} \exp \left[ - \sum_{i,j} \hat{a}_i^\dag G_{i,j} \hat{a}_j \right],
\eeq
where the matrix $G$ is known as the `fictitious Hamiltonian' relating with the Hamiltonian of the system as $G=\beta H$. Due to the translational invariance, $G$ is diagonal in the Bloch basis, namely \cite{Bardyn_PRX_2018}:
\beq
G = \sum_{\bk} G_{\bk} |\bk\rangle \langle \bk |,
\eeq
where $|\bk\rangle$ denotes the Bloch basis. $G_{\bk}$ is a non-diagonal Hermitian matrix defined in band space (denoted by index $s$), with elements $[G_{\bk}]_{s,s'}$, and can be diagonalized by a unitary transformation:
\beq \label{EGP_diagonal}
B_\bk = \mathrm{diag}_s(\beta_{\bk,s}) = \mathcal{U}_\bk^\dag G_\bk \mathcal{U}_{\bk},
\eeq
where for the Hamiltonian with $\mathcal{N}$ energy bands $\mathcal{U}_{\bk}$ is constructed from the $n$ eigenvectors as:
\beq \label{def-U}
\mathcal{U}_{\bk} = \left( \Psi_1(\bk), \Psi_2(\bk), \cdots , \Psi_\mathcal{N}(\bk) \right).
\eeq
Eq.~(\ref{EGP_diagonal}) explicitly indicates that the two important quantities in the calculations of ensemble geometric phase $B_\bk$ and $\mathcal{U}_\bk$ are related to the coefficient matrix $G_{\bk}$ in the Bloch basis \cite{Bardyn_PRX_2018}, namely $G_{\bk}$ diagonalized into $B_\bk$ by the unitary matrix $\mathcal{U}_\bk$.

The straightforward calculation shows that the ensemble geometry phase can be expressed in terms of $G_\bk$, $B_{\bk}$ and $\mathcal{U}_{\bk}$:
\beq \label{EGP}
\varphi_E(k_y) = \Im m \left[\ln \mathrm{det} \left( 1+M_T \right) \right],
\eeq
where:
\beq \label{def_MT}
M_T = (-1)^{N_L+1} \prod_{\bk \in \mathcal{L}(k_y)} e^{-B_\bk} \mathcal{U}_{\bk+\delta k_x \hat{x}}^\dag  \mathcal{U}_{\bk}.
\eeq
Here $N_L$ is the number of grid points along the loop $\mathcal{L}(k_y)$. In the evaluation of $M_T$, the important component is the product $\mathcal{U}_{\bk+\delta k_x \hat{x}}^\dag  \mathcal{U}_{\bk}$. From the definition of $\mathcal{U}_{\bk}$ given by Eq.~(\ref{def-U}), we can easily identify that $\mathcal{U}_{\bk+\delta k_x \hat{x}}^\dag  \mathcal{U}_{\bk}$ is the generalization of the normalized overlap, which includes the parallel transports not only of the intraband wavefunction but also of the interband one. 

\begin{figure*}[t]
  \centering
  \includegraphics[width=1.8\columnwidth]{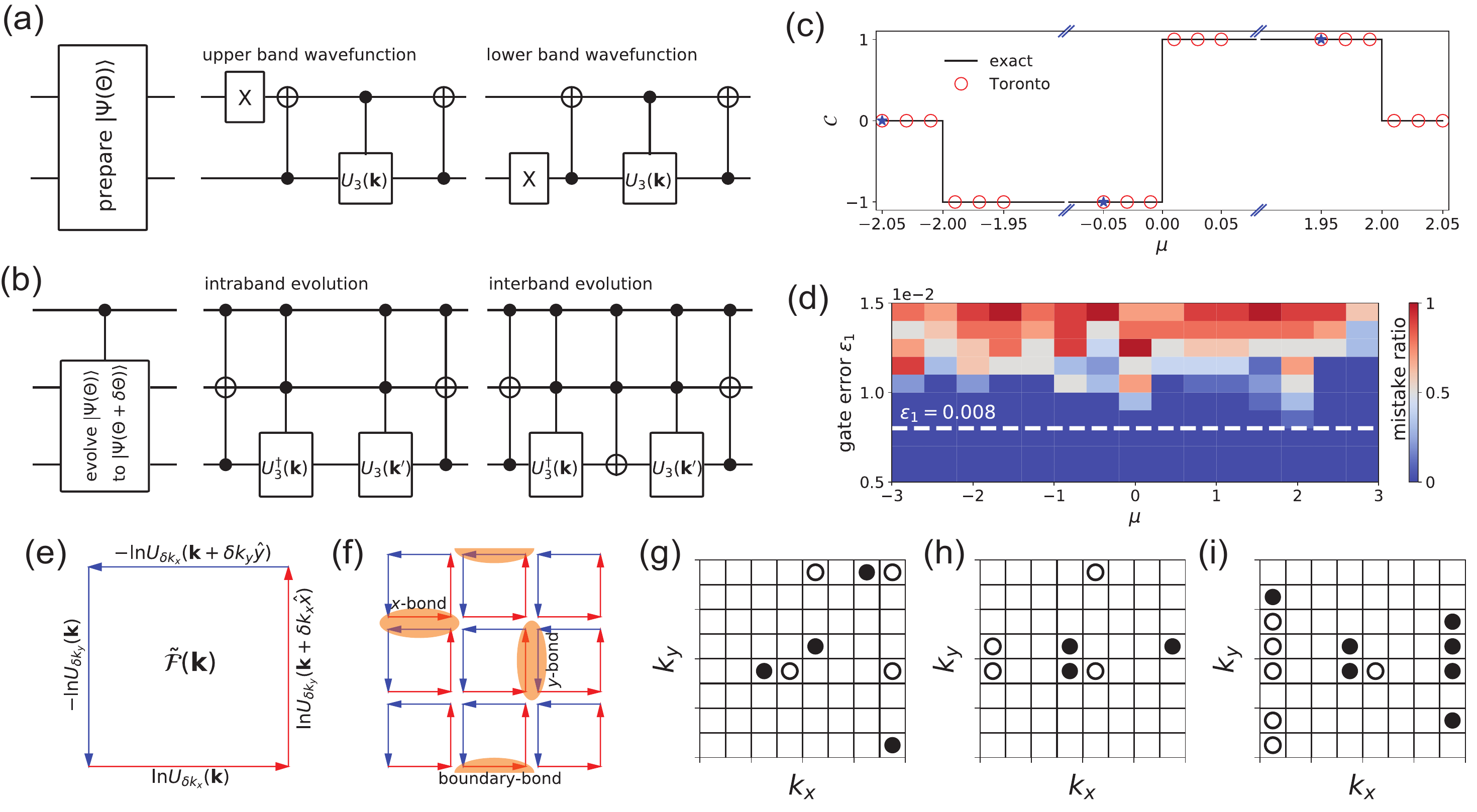}
  \caption{Robust measurement of Chern numbers of chiral $p$-wave superconductors: (a) Detailed circuit to perform single-particle wave
function preparation; (b) Detailed circuit to perform wave function evolution; In the panels (a) and (b), $U_3(\bk)$ ($U_3^\dag(\bk)$) is a general single-qubit operator specified by three angles determined at the momentum point $\bk$ for a given $\mu$, `X' is a Pauli-X gate, 
$\bullet$ denotes a control, and $\oplus$ denotes a NOT gate. (c) Topological phase diagram for chiral $p$-wave superconductors determined by the measured Chern numbers on IBMQ-Toronto; (d) Mistake ratio of Chern number measurements from noisy simulations by assuming that the two-qubit gate error $\varepsilon_2$ is $10$x the one-qubit gate error $\varepsilon_1$; (e) The structure of local gauge field $\tilde{F}(\bk)$ defined by Eq.~(\ref{Eq4}); The red arrows correspond to plus signs in front of the terms, while the blue arrows correspond to the minus signs. (f) the summation of $\tilde{F}(\bk)$ over the BZ leads to a perfect cancellation: the orange highlighted parts show the cancellations happening at an $x$, a $y$ and a boundary bond; (g)-(i) Integer-valued field $n(\bk)$ extracted from the data points denoted by blue stars in (c) ($\bullet$ denotes $n=1$, $\circ$ denotes $n=-1$ and the empty box is for $n=0$). The overlap $U_{\delta \bk}(\bk)$ measured by both IBMQ-Toronto and noisy simulations was obtained with $N=5120$ shots.}
   \label{fig2}
\end{figure*}

\section{Exactly solvable model: chiral $p$-wave superconductor}
\label{sec3}
For our first example, we consider a model for two-dimensional chiral $p$-wave superconductors \cite{Volovik_JEPT_1999,Read_PRB_2000}, which can be tuned through several trivial and topological phases. This model has the Hamiltonian density at momentum point $\bk$
\begin{align} \label{Hamiltonian}
\mathcal{H}(\bk) = 	&	\Delta (\sin k_y \sigma_x + \sin k_x \sigma_y) - \mathcal{E}(\bk) \sigma_z,
\end{align}
where $\mathcal{E}(\bk) = t ( \cos k_x + \cos k_y ) + \mu$ with $t$ and $\mu$ denoting the hopping strength and the chemical potential in the normal state, respectively, and $\Delta$ is the superconducting gap; the Hamiltonian is the integral of this density over all $\bk$. We set $t=\Delta=1$ so that the different phases are tuned by $\mu$ only. The topological quantum critical points occur when the energy levels of the two bands touch at some point in $\bk$-space. As shown in Fig.~\ref{fig2}, this model exhibits $4$ different phases separated by $3$ topological quantum critical points at $\mu_c=\{-2,0,2\}$. The trivial phases with $\mathcal{C}=0$ occur for $|\mu|>2$; when $|\mu|<2$, $\mathcal{C}=\mathrm{sign}(\mu)$.

To measure the topological invariants of this model, we need to explicitly construct the quantum circuit to measure the normalized overlap $U_{\delta \bk}(\bk)$ and its generalization $\mathcal{U}_{\bk+\delta k_x \hat{x}}^\dag  \mathcal{U}_{\bk}$. Because this model can be solved exactly, we can construct the exact circuit. In the following, we outline how to measure the normalized overlap of a wave function at neighboring momentum-space mesh points by the Hadamard test.

\subsection{Construction of the exact circuit to measure the normalized overlap}

First we denote the prepared state as $|\Psi\rangle=|0\rangle \otimes |\psi\rangle$, where $|0\rangle$ is the initial state of the ancilla and $|\psi\rangle$ is the wave function at one of the mesh points in the BZ. We apply the Hardmard gate to the ancilla, resulting in the following product state: 
\beq
|\Psi\rangle = \frac{1}{\sqrt{2}} |0\rangle \otimes |\psi\rangle + \frac{1}{\sqrt{2}} |1\rangle \otimes |\psi\rangle.
\eeq
Then, we apply the controlled $\mathfrak{U}$ operation, where $\mathfrak{U}$ relates the wave functions at the neighboring mesh points. After applying the operation, the state is in an entangled superposition given by
\beq
|\Psi\rangle = \frac{1}{\sqrt{2}} |0\rangle \otimes |\psi\rangle + \frac{1}{\sqrt{2}} |1\rangle \otimes \mathfrak{U} |\psi\rangle.
\eeq
Finally, we projectively measure the expectation values of $\sigma_x \otimes I$ and $\sigma_y\otimes I$, which  give the real and imaginary parts of the overlap:
\beq
\langle \Psi | \sigma_x \otimes I | \Psi \rangle = \Re e \langle \psi | \mathfrak{U} |\psi\rangle,
\eeq
and
\beq
\langle \Psi | \sigma_y \otimes I | \Psi \rangle = \Im m \langle \psi | \mathfrak{U} |\psi\rangle.
\eeq

To complete the description of the algorithm, we have two more steps: first, we need to determine what the initial state $|\psi\rangle$ is that we will use and how we prepare it by a quantum circuit and second, we need to determine the unitary operator $\mathfrak{U}$ that evolves the wave function between neighboring mesh points and how we can realize it as a quantum circuit. We answer these two questions next.

\subsubsection{Preparation of the initial wave function}

We begin from the Hamiltonian density given by Eq.~(\ref{Hamiltonian}). The full Hamiltonian can be written as:
\beq
H = \sum_{\bk} \left( \begin{array}{cc} c_{\bk}^\dag & c_{\bk} \end{array}\right) \mathcal{H}(\bk) \left( \begin{array}{c} c_{\bk} \\ c_{\bk}^\dag \end{array}\right),
\eeq
where $\mathcal{H}(\bk)$ is given by Eq.~(\ref{Hamiltonian}), and the BZ is defined by $k_x \in [-\pi,\pi]$ and $k_y\in[-\pi,\pi]$. $\mathcal{H}(\bk)$ has the eigenvalues:
\beq
E_{\pm} = \pm\sqrt{\left(\sin^2 k_y + \sin^2 k_x \right)+ \left[ (\cos k_x + \cos k_y) + \mu \right]^2 },
\eeq
where we used explicitly $\Delta=t=1$. We find that the gap between the two energy bands closes at $(k_x=0,k_y=0)$ for $\mu=-2$, at $(k_x=\pm \pi, k_y=\pm \pi)$ for $\mu=2$, and at $(k_x=\pm\pi,k_y=\mp \pi)$ for $\mu=0$. These gap closing points separate different topological phases.

We define angles $\theta(\bk)$ and $\varphi(\bk)$, determined at each momentum point, via:
\beq \label{theta}
\cos \theta = \frac{(\cos k_x + \cos k_y) + \mu}{E_{+}},
\eeq
and
\beq \label{phi}
\cos \varphi = \frac{ \sin k_y}{\sqrt{ \left(\sin^2 k_y + \sin^2 k_x \right)}},
\eeq
so that the corresponding eigenvectors of $E_{\pm}$ can be written as:
\beq
\Psi_{+}(\bk) = \left( \begin{array}{c} \cos \frac{\theta}{2} \\ \phantom{-}\sin \frac{\theta}{2} e^{-i\varphi} \end{array}\right),~\Psi_{-}(\bk) = \left( \begin{array}{c} -\sin \frac{\theta}{2} e^{i\varphi} \\ \cos \frac{\theta}{2}  \end{array}\right).
\eeq
This eigensolution indicates that:
\beq
\mathrm{diag}\left( E_+(\bk), E_-(\bk) \right) = \mathcal{U}^\dag(\bk) \mathcal{H}(\bk) \mathcal{U}(\bk),
\eeq
where $\mathcal{U}(\bk)$ is a special case of Eq.~(\ref{def-U}):
\beq
\mathcal{U}(\bk) = \left(\begin{array}{cc} \Psi_+(\bk) & \Psi_-(\bk) \end{array}\right)
\eeq
We denote this diagonal representation formed by the energy eigenstates of $\mathcal{H}(\bk)$ as the band representation, and the corresponding annihilation operators for the $E_+$ and $E_-$ bands are denoted by $f_{\bk}$ and  $f_{\bk}^\dag$. They are related to the original $c_{\bk}$ and $c_{\bk}^\dag$ via
\beq \label{relation}
\left( \begin{array}{c} f_{\bk} \\ f_{\bk}^\dag \end{array}\right) = \mathcal{U}^\dag(\bk) \left( \begin{array}{c} c_{\bk} \\ c_{\bk}^\dag \end{array}\right).
\eeq

The topological invariant is calculated from the wave function in the Brillouin zone. We begin from the diagonalized band representation: the initial states are either $|0_f 1_{f^\dag}\rangle$ for the lower band or $|1_f 0_{f^\dag}\rangle$ for the upper band. The wave function can be constructed by the operators $c_{\bk}^\dag$ and  $c_{\bk}$ acting on the vacuum. The relation between $(f_{\bk}, f_{\bk}^\dag)$ and $(c_{\bk}, c_{\bk}^\dag)$ is clear from Eq.~(\ref{relation}), or more explicitly 
\beq
\begin{cases}
c_{\bk}^\dag = \cos \frac{\theta}{2} f_{\bk}^\dag - \sin \frac{\theta}{2} e^{-i\varphi} f_{\bk}, \\
c_{\bk} = \sin \frac{\theta}{2} e^{i\varphi} f_{\bk}^\dag + \cos \frac{\theta}{2} f_{\bk}.
\end{cases}
\eeq
Hence, the following relation can be found that relates the two representations:
\begin{align}
\left(\begin{array}{c} |0_{c} 0_{c^\dag}\rangle \\ |1_{c} 0_{c^\dag}\rangle \\ |0_{c} 1_{c^\dag}\rangle \\ |1_{c} 1_{c^\dag}\rangle \end{array}\right) = V(\theta,\varphi) \left(\begin{array}{c} |0_{f} 0_{f^\dag}\rangle \\ |1_{f} 0_{f^\dag}\rangle \\ |0_{f} 1_{f^\dag}\rangle \\ |1_{f} 1_{f^\dag}\rangle \end{array}\right),
\end{align}
where the transformation matrix $V$ is given by:
\beq
V(\theta,\varphi) = \left(\begin{array}{cccc} 1 & 0 & 0 & 0 \\ 0 & \cos \frac{\theta}{2} & -\sin \frac{\theta}{2} e^{-i\varphi} & 0 \\ 0 & \sin \frac{\theta}{2} e^{i\varphi} & \cos \frac{\theta}{2} & 0 \\ 0 & 0 & 0 & 1 \end{array}\right).
\eeq
From this relation, we know that the state $|\psi\rangle$ is obtained by applying $V$ on either $|1_f 0_{f^\dag}\rangle$ for the upper band or $|0_f 1_{f^\dag}\rangle$ for the lower band. In the language of QISKIT \cite{qiskit2019}, this operation can be realized by two \textsf{CNOT} gates and a controlled-$U_3$ gate:
\begin{align}
V(\theta,\varphi) &= \mathrm{CNOT}[q_1,q_0] \nonumber \\
 &\times\mathrm{CU}_3[q_0,q_1](\vartheta=\theta,\lambda = -\varphi, \phi=\varphi) \nonumber \\
 &\times \mathrm{CNOT}[q_1,q_0],
\end{align}
where the first qubit in the bracket is the control qubit and the second one is the target qubit. The matrix form of $\mathrm{CU}_3[q_0,q_1](\vartheta,\lambda, \phi)$ is:
\beq \label{CU}
\mathrm{CU}_3[q_0,q_1](\vartheta,\lambda, \phi) = \left(\begin{array}{cccc} 1 & 0 & 0 & 0 \\ 0 & \cos \frac{\vartheta}{2} & 0 & -\sin \frac{\vartheta}{2} e^{i\lambda} \\ 0 & 0 & 1 & 0 \\ 0 & e^{i\phi} \sin \frac{\vartheta}{2} & 0 & e^{i(\lambda+\phi)} \cos \frac{\vartheta}{2} \end{array}\right).
\eeq
$\mathrm{CU}_3[q_0,q_1](\vartheta,\lambda, \phi)$ is realized by a general single-qubit operator $U_3$ under the control of an ancilla. In particular, the operator $U_3$, appearing in Fig.~\ref{fig2}, can be written in terms of $\theta(\bk)$ and $\varphi(\bk)$ defined by Eq.~(\ref{theta}) and (\ref{phi}):
\beq
U_3(\vartheta=\theta,\lambda = -\varphi, \phi=\varphi) = \left(\begin{array}{cc} \cos \frac{\vartheta}{2} & -\sin \frac{\vartheta}{2} e^{i\lambda} \\ e^{i\phi} \sin \frac{\vartheta}{2} & e^{i(\lambda+\phi)} \cos\frac{\vartheta}{2} \end{array}\right).
\eeq

\begin{figure}[!h]
  \centering
  \includegraphics[width=0.8\columnwidth]{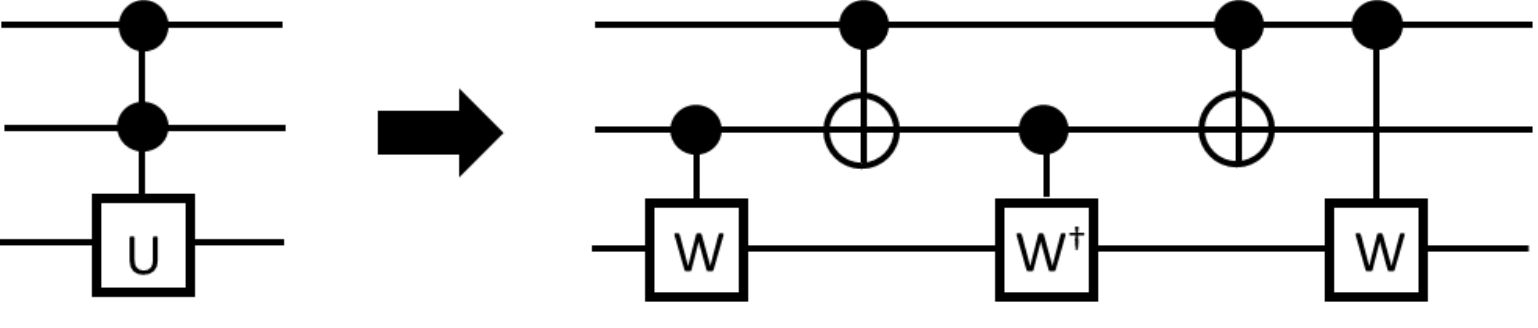}
  \caption{The realization of the controlled-controlled-unitary gate by two-qubit gates, where $W$ fulfills the condition: $W^2 = U$.}
   \label{fig3}
\end{figure}

\subsubsection{Relating wave functions at neighboring mesh points}

From the discussion in the last subsection, we see that the state at a particular point $\bk$ in the BZ can be prepared by $V(\theta_{\bk},\varphi_{\bk})$. Therefore, the operation transforming the wave function from $\bk$ to  $\bk'$ is
\beq
\mathcal{V}(\bk';\bk) = V(\theta_{\bk'},\varphi_{\bk'}) V^\dag(\theta_{\bk},\varphi_{\bk}).
\eeq
To measure the overlap of the wave function, we use the Hadamard test with an ancilla qubit controlling the application of $\mathcal{V}(\bk';\bk)$ on the other two qubits. This means that the equation above is modified to
\beq
\mathrm{C}\mathcal{V}(\bk';\bk) = \mathrm{CV}(\theta_{\bk'},\varphi_{\bk'}) \mathrm{CV}^\dag(\theta_{\bk},\varphi_{\bk}).
\eeq
The extra letter $\mathrm{C}$ indicates that the two-qubit unitary operation is controlled by an ancilla qubit. For each $\mathrm{CV}$ operation, we can realize it by extending the two-qubit gate in Eq.~(\ref{CU}) as follows:
\begin{align}
\mathrm{CV}(\theta,\varphi) &= \mathrm{CCX}[q_0,q_2,q_1] \nonumber \\
&\times\mathrm{CCU}_3[q_0,q_1,q_2](\vartheta=\theta,\lambda = -\varphi, \phi=\varphi) \nonumber \\
&\times\mathrm{CCX}[q_0,q_2,q_1],
\end{align}
where the first two qubits in the bracket are the controlled qubits and the last one is the target qubit. The $\mathrm{CCX}$ is the well-known Toffoli gate, so we just need to construct the $\mathrm{CCU}_3$ gate. This can be accomplished by the circuit shown in Fig.~\ref{fig3}. Using these components, the quantum circuits for the chiral $p$-wave superconductors shown in Fig.~2(a) and (b) in the main text can be completed. Using these components, the quantum circuit to measure each overlap for the chiral $p$-wave superconducting model has a depth of $25$ including $58$ \textsf{CNOT} gates on IBMQ machines.

\subsection{Calculating the Chern number}
\label{sec:calc_chern}
Following the detailed construction procedures stated in the last subsection, the general circuit given in Fig.~\ref{fig1}(c) with the wave function preparation and evolution components is explicitly realized as shown in Fig.~\ref{fig2}(a) and (b). In our measurements we use a uniform discretization of the BZ into $8\times8$ mesh points, beyond the minimal discretization constrained by the admissibility condition \cite{Fukui_JPSJ_2005}.  After measuring the normalized overlap $U_{\delta \bk}(\bk)$ associated with each bond connecting neighboring mesh points in the BZ, the Chern number can be extracted by using Eq.~(\ref{Eq1}). We first demonstrate the measurement of the Chern number on the IBMQ-Toronto machine, focusing on the regions near the topological critical points, which are typically most sensitive to noise. Fig.~\ref{fig2}(c) shows the exact results for the Chern number $\mathcal{C}$ as a function of $\mu$ in the black curves and the results from IBMQ-Toronto as circles. Remarkably, we observe that the Chern number measured on the quantum computer is {\it error-free}.

The robustness of the measurement of Chern numbers on quantum hardware can be understood by introducing another local gauge field $\tilde{\mathcal{F}}(\bk)$:
\begin{align}\label{Eq4}
\tilde{\mathcal{F}}(\bk) &= \left[\ln U_{\delta \hat{\bk}_x}(\bk) - \ln U_{\delta \hat{\bk}_x}(\bk+\delta \hat{\bk}_y)\right] \nonumber \\
&+ \left[\ln U_{\delta \hat{\bk}_y}(\bk+\delta \hat{\bk}_x ) - \ln  U_{\delta \hat{\bk}_y}(\bk)\right].
\end{align}
It relates to $\mathcal{F}(\bk)$ defined in Eq.(\ref{Eq2}) as $\mathcal{F}(\bk) = \tilde{\mathcal{F}}(\bk) + i2\pi n(\bk)$, where $n(\bk)$ denotes an integer-value field to guarantee $\mathcal{F}(\bk) \in [-\pi,\pi]$. The structure of $\tilde{\mathcal{F}}(\bk)$ defined on the plaquette formed by the neighboring mesh points is shown in Fig.~\ref{fig2}(e). The logarithm of the normalized overlap associated to the bond shared by two neighboring plaquettes will contribute oppositely to $\tilde{\mathcal{F}}$ defined on the two plaquettes. Thus, when we sum $\tilde{\mathcal{F}}$ over the BZ, a perfect cancellation occurs as illustrated in Fig.~\ref{fig2}(f), which leads to:
\beq \label{chern_integer}
\mathcal{C} = \sum_{\bk} n(\bk).
\eeq 
The perfect cancellation of $\tilde{\mathcal{F}}$ over the BZ indicates that the individual errors of $U_{\delta \bk}(\bk)$ will always be removed upon the summation, which means that the measurement is entirely immune to the separated local noise. The integer-value field $n(\bk)$ can be extracted correctly in each plaquette, as long as $\tilde{\mathcal{F}}$ can be measured with an affordable error smaller than $2\pi$, which provides another protection of the measurements on quantum hardware. As a self-consistent check, we extract $n(\bk)$ for three typical $\mu$ values (denoted by the blue stars in Fig.~\ref{fig2}(c)), and show them in Fig.~\ref{fig2}(g)-(i). The summation of $n(k)$ is indeed consistent with the measured Chern number shown in Fig.~\ref{fig2}(c). 

Moreover, The relationship between the two local gauge fields $\mathcal{F}(\bk)$ and $\tilde{\mathcal{F}}(\bk)$ indicates that the expression of Chern numbers given by Eq.~(\ref{Eq1}) is identical to Eq.~(\ref{chern_integer}). This identity guarantees that Chern numbers calculated from the measured wave function overlaps by quantum circuits proposed in Fig.~\ref{fig1} are integers by definition, {\it namely} the sum of integers is still an integer.

Of course, if the hardware noise gets sufficiently large, this approach must break down. To investigate how the measurement result is affected by the machine noise, we perform noisy simulations on classical computer with depolarizing errors \cite{Knill_Nature_2005,Cross_QIC_2009} introduced to the single-qubit and two-qubit gates (see Appendix \ref{appendix_a} for more details). For IBM machines, the typical two-qubit gate error $\varepsilon_2$ is about one order of magnitude larger than the single-qubit gate error $\varepsilon_1$. In our simulations, we set $\varepsilon_2=10\varepsilon_1$ and successively tune $\varepsilon_1$ from $\varepsilon_1=0.005$ to $0.015$ with a step size of $0.001$. For each $\varepsilon_1$, $10$ trials were performed. We define the mistake ratio of the measurements as the percentage of incorrect $\mathcal{C}$ values. The results of the noisy simulations are shown in Fig.~\ref{fig2}(d). As expected, larger gate error leads to larger mistake ratios; however, the measured Chern number is error-free when $\varepsilon_1\leq0.008$. 

Here we need to emphasize that the threshold $\varepsilon_1$ is determined with only $10$ trials, so it is not very accurate. A better way to determine the threshold is to gradually increase the trial numbers to find a saturated threshold value. In Appendix \ref{appendix_b}, we have done such analysis and found that the threshold saturates at $\varepsilon_1=0.006$. However, this result does not affect the above discussion.

In our strategy, the integer field $n(\bk)$ is extracted from the the difference between $F(\bk)$ and $\tilde{F}(\bk)$. The counting statistic errors, such as shot noises, are not guaranteed to be eliminated. One might expect that the influence of shot noise errors would affect the accuracy of the calculation of Chern numbers, when the system is close to a topological critical point, where the admissibility is much easier to break. However, our extensive noisy simulations demonstrated that even when the system is very close to a topological critical point, {\it i.e.} $|\mu-\mu_c|=0.0001$, the threshold $\varepsilon_1$ determined from noisy simulations does not change (see Appendix \ref{appendix_b1} for more details). Therefore, the noisy simulations indicates that shot noise does not play any practical roles in correctly calculating Chern numbers with the strategies proposed here.

\subsection{Measuring the winding of the Zak phase}

\begin{figure}[h]
  \centering
  \includegraphics[width=0.9\columnwidth]{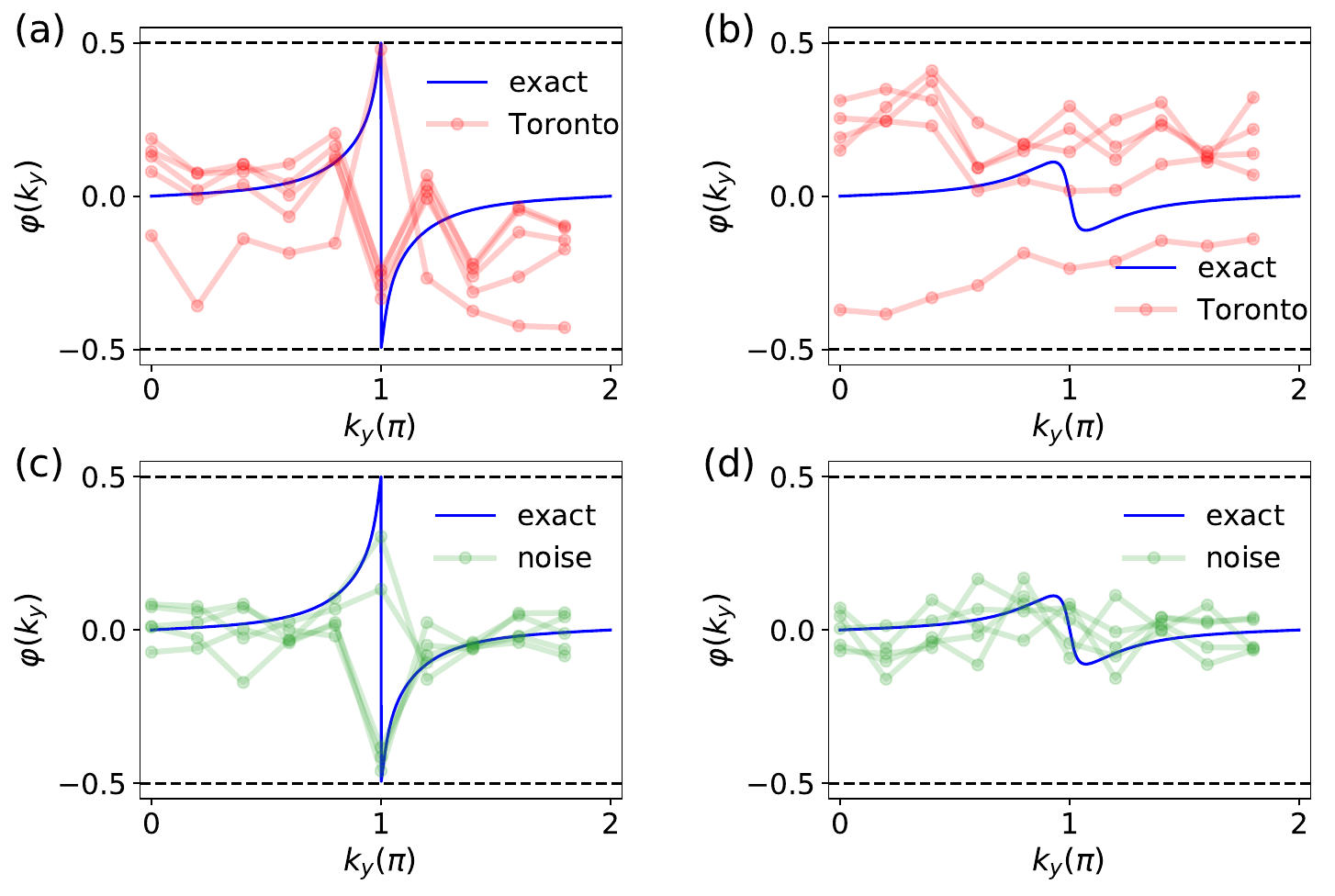}
  \caption{Qualitative robustness of Zak phase on NISQ machines: the winding of Zak phase $\varphi(k_y)$ along $k_y$ measured by IBMQ-Toronto: (a) topological phase with $\mu=1.9$; (b) trivial phase with $\mu=2.1$. For the comparison, the winding of the Zak phase was measured by noise simulations and shown in (c) for $\mu=1.9$ and (d) for $\mu=2.1$. In the noise simulations, the gate noises were chosen to be $\epsilon_1=0.008$ and $\epsilon_2=0.08$. Each overlap $U_{\delta \bk}(\bk)$ measured by IBMQ-Toronto and noise simulations were obtained by $N=5120$ shots, and $N_L=8$ grids point were used for each loop $\mathcal{L}(k_y)$ to obtain Zak phase.}
   \label{fig4}
\end{figure}

Next, we illustrate the calculation of the Zak phase, which does not enjoy a similar level of robustness.
To measure the Zak phase associated to a particular $k_y$, we need to measure the normalized overlap $U_{\delta \hat{\bk}_x}(\bk)$ along the mesh points along the $x$-direction by using the quantum circuit shown in Fig.~\ref{fig2}(a) and (b). Then through Eq.~(\ref{Zak}) the Zak phase for the particular $k_y$ can be obtained. In Fig.~\ref{fig4}(a) and (b), the Zak phases $\varphi(k_y)$ obtained from $5$ independent simulations on IBMQ-Toronto were plotted as functions of $k_y$ for a typical topological state (with $\mu=1.9$) and a typical trivial state ($\mu=2.1$) respectively. Here the results from quantum machines do not fall exactly on the exact results (the blue curves), but the results from quantum computers do capture the main features of the two topologically distinct phases. In particular, for the topological state, a sharp change at the high-symmetry point $k_y=\pi$ (see Fig.~\ref{fig4}(a)) is identified and signifies the non-trivial winding of the Zak phase along $k_y$, while this sharp change is absent for the trivial phase as illustrated in Fig.~\ref{fig4}(b). For the comparison, we also performed noise simulations by setting $\epsilon_1=0.008$ and $\epsilon_2=10\epsilon_1$ for the two states studied in Fig.~\ref{fig4}(a) and (b). The results for $5$ independent simulations for the topological state with $\mu=1.9$ were plotted in Fig.~\ref{fig4}(c), and the results for the trivial state with $\mu=2.1$ were in Fig.~\ref{fig4}(d). It can be clearly observed that the noise simulation results are quite similar with those from IBMQ-Toronto. These results from both IBMQ-Toronto and noise simulations suggest that the topology of wavefunctions can be successfully identified by measuring the winding of Zak phases; however, the error cancellation mechanism that occurs for the Chern number no longer applies here. 

\subsection{Measuring the ensemble geometry phase}

\begin{figure}[h]
  \centering
  \includegraphics[width=0.9\columnwidth]{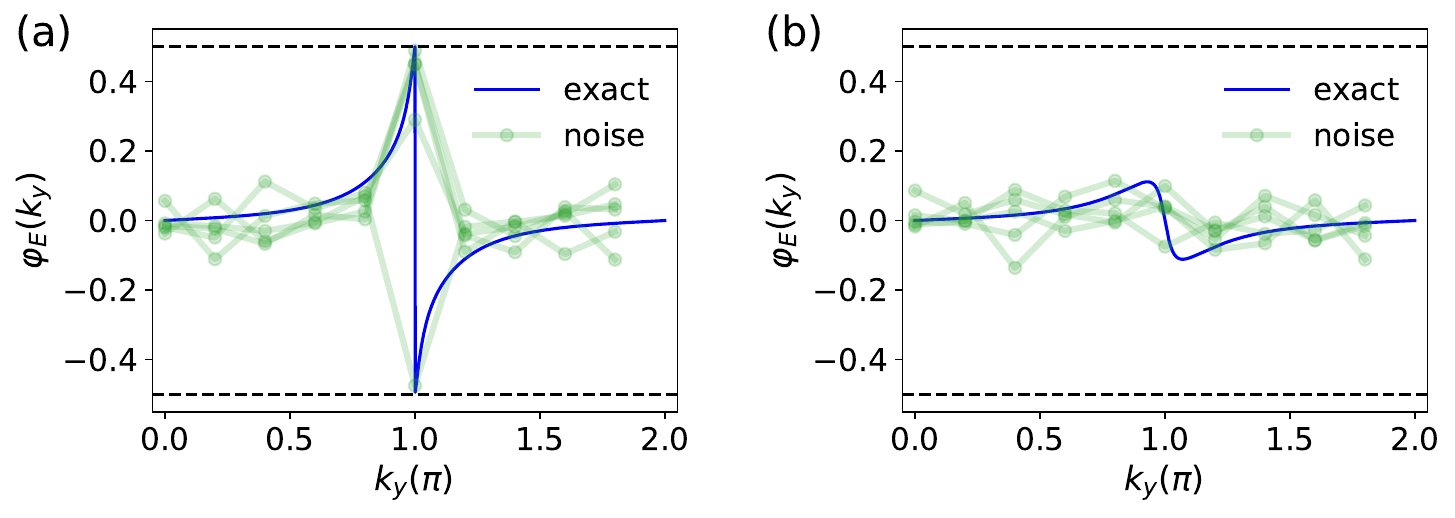}
  \caption{Qualitative robustness of ensemble geometric phase illustrated by noise simulations: (a) the winding of ensemble geometric phase $\varphi_E(k_y)$ along $k_y$ for (a) topological phase with $\mu=1.9$ and (b) trivial phase with $\mu=2.1$. The number of grid points for the loop $\mathcal{L}(k_y)$ is $N_L=8$ and the inverse temperature $\beta=2.1$. In this simulation, the gate noises were chosen to be $\epsilon_1=0.008$ and $\epsilon_2=0.08$, and $N=5120$ shots were used to obtain each element of $\mathcal{U}_{\bk+\delta k_x \hat{x}}^\dag  \mathcal{U}_{\bk}$.}
   \label{fig5}
\end{figure}

As we have emphasized in the general discussion provided in Sec. \ref{sec2}, the measurement of the ensemble geometric phase requires us to not only measure the intraband normalized overlaps but also the interband normalized overlaps (see Appendix \ref{appendix_c} for details). Once we obtain these normalized overlaps, we can calculate the ensemble geometric phase from Eq.~(\ref{EGP}). We demonstrate the measurement of ensemble geometric phase by noise simulations, which, as we show in Fig.~\ref{fig2} and Fig.~\ref{fig4}, can have faithful results compatible with those from real quantum hardware. The winding of the ensemble geometric phase for the topological phase with $\mu=1.9$ and for the trivial phase with $\mu=2.1$ is shown in Fig.~\ref{fig5}(a) and (b) respectively. Indeed the results are similar to the that of Zak phase; and as we have argued in the above, the two phases with different topology can be qualitatively distinguished.

\section{Variationally prepared states} \label{sec4}

For generic (interacting) quantum states, finding the exact circuits to prepare and evolve the wave function is a difficult task. However, given the robustness of the topological invariant, it is possible to replace the exact circuits (or wave functions) by approximate ones. Here we use  adaptive VQE \cite{Grimsley_NC_2019} to approximate the wave function preparation and evolution circuits. With this technique combined with our strategy, the topological invariant for arbitrary models, even \emph{interacting ones}, can be calculated with quantum hardware. Considering that VQE methods in principle can be scaled to calculate large systems beyond the capability of classical computers, we expect that our strategy with VQE can allow for the calculation of topological invariants of these more complex models. 

We first demonstrate our strategy by calculating the Chern number for a two-particle state in a prototype topological model, the flux $2\pi/3$ quantum Hall model. As above, the Chern numbers can be calculated accurately on the present quantum hardware without any error. Then we introduce a Hatsugai-Kohmoto interaction, whose properties are actively studied recently in strongly correlated models \cite{Hatsugai_JPSJ_1992,Philips_NP_2020} demonstrating that our strategy even works when the system is interacting. Our results show that the Chern number for the interacting model can still be accurately calculated on IBM's quantum hardware. 

Finally, to further demonstrate the generality of our strategy, we applied it to calculate Chern numbers for interacting models on a real-space lattice. In this case, the calculation is be performed in a finite-size cluster in real space with twist-angle boundary conditions. The calculations of Chern numbers for a $4$-site interacting Chern insulator is demonstrated with noisy quantum simulators \cite{qiskit2019}.

\subsection{Calculating Chern numbers for a two-particle Quantum Hall State} \label{sec4a}

We present the calculation of Chern numbers for the two-particle ground state of the flux-$2\pi/3$ fermionic quantum Hall model. The other topological invariants can be also straightforwardly measured by following the same procedures provided in Sec.~\ref{sec3}, and we do not show them here. 

After choosing the hopping strength as the energy unit, the Hamiltonian of the system is given by \cite{Fukui_JPSJ_2005}: 
\beq
H = - \sum_{x,y} \left( c_{x+1,y}^\dag c_{x,y} + e^{-i\Phi x}  c_{x,y+1}^\dag c_{x,y} \right) + h.c.,
\eeq
where $c_{x,y}$ ($c_{x,y}^\dag$) is the fermionic annihilation (creation) operator on site $(x,y)$. The magnetic flux per plaquette is set to be $\Phi = 2\pi/3$ so that this model can be simulated with $3$ qubits and the magnetic BZ is defined with $k_x \in [0,2\pi/3]$ and $k_y \in [0,2\pi]$. For this particular model, the ground state wave function can be prepared by a generalized factorized unitary coupled cluster ansatz that is truncated at single-particle excitations/de-excitations \cite{Grimsley_NC_2019,Lee_JCTC_2019,Chen_JCTC_2021}. The operator pool for the adaptive VQE \cite{Grimsley_NC_2019} simply includes all possible operators generating  single excitations.

\begin{figure}[t]
  \centering
  \includegraphics[width=1\columnwidth]{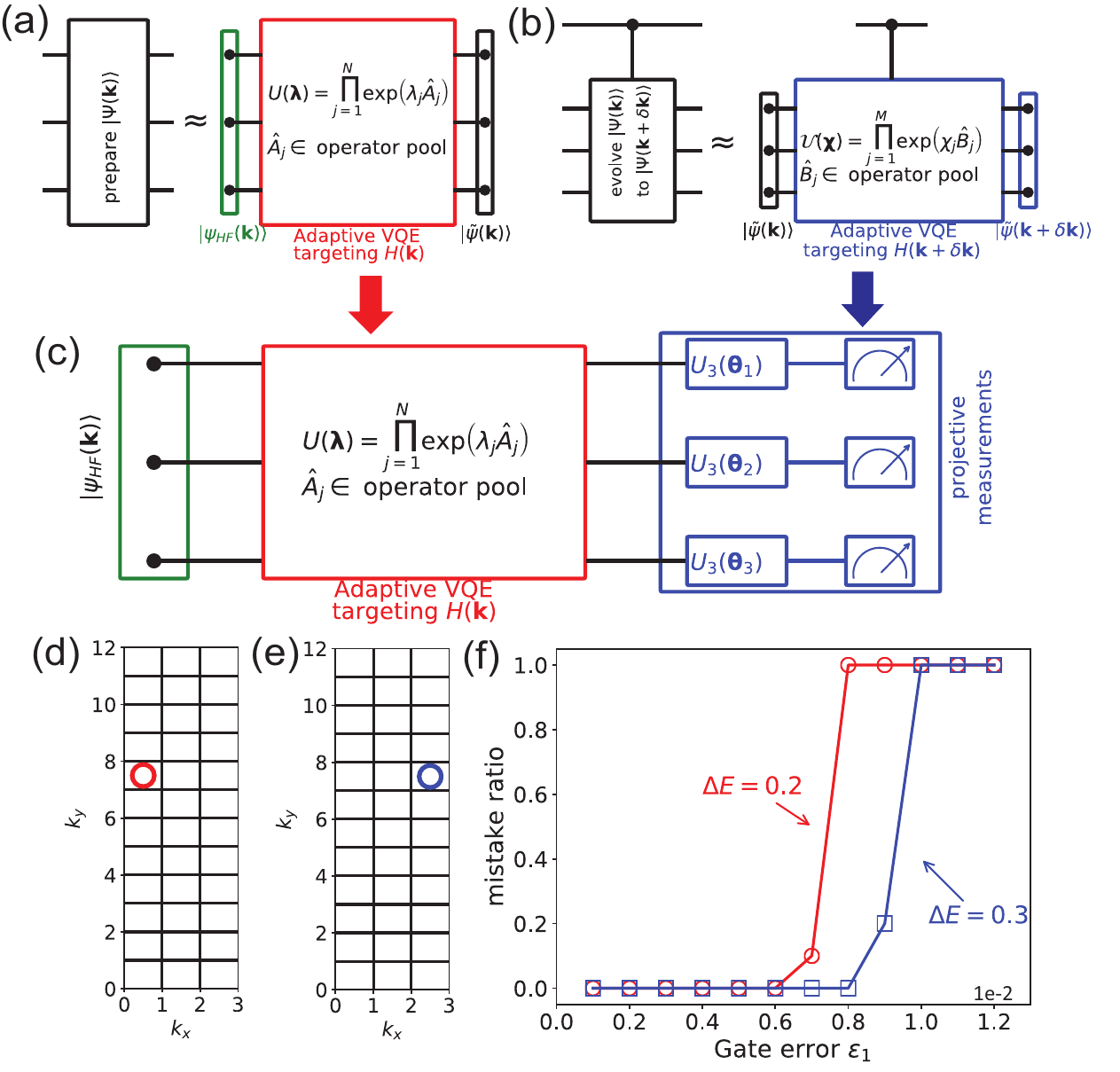}
  \caption{Robust measurement of the Chern number via adaptive VQE for the two-body ground state of the flux-$2\pi/3$ quantum Hall model: (a) the approximated circuit to prepare the wave function at $\bk$ via adaptive VQE; (b) the approximated circuit generated by adaptive VQE to evolve the wave function from $\bk$ to $\bk+\delta \bk$; (c) the approximate circuit to measure the normalized overlap $U_{\delta \bk}(\bk)$ via adaptive VQE. In this implementation, the circuit to evolve the wave function
  is done by direct projective measurements. The calculated integer-valued field $n(\bk)$ in the discretized magnetic BZ on IBMQ-Montreal are shown in (d) and (e) for $\Delta E =0.2$ and $\Delta E =0.3$ respectively; (f) Mistake ratio of Chern number measurements from noisy simulations versus the single-qubit gate error $\varepsilon_1$ for $\Delta = 0.2$ and $\Delta = 0.3$, respectively. In all cases, $N=8196$ shots were used to obtain the normalized overlap $U_{\delta \bk}(\bk)$.}
   \label{fig6}
\end{figure}

To calculate the Chern number we first transform the Hamiltonian into the momentum space by introducing a magnetic BZ:
\beq
H = \sum_{\bk} \Psi_{\bk}^\dag \mathcal{H}(\bk) \Psi_{\bk},
\eeq
where $\Psi_{\bk} = \left( c_{\bk;1}, c_{\bk;2}, c_{\bk;3} \right)^T$ is the annihilation operator for the fermions at \emph{the three effective orbitals}, and the Hamiltonian density is
\begin{align} \label{Hmatrix}
\mathcal{H}(\bk) 
=-\left(\begin{array}{ccc} 2 \cos k_y & 1 & e^{-i3k_x} \\ 1 & 2 \cos(k_y+\frac{2\pi}{3}) & 1 \\ e^{i3k_x} & 1 & 2\cos(k_y+\frac{4\pi}{3}) \end{array}\right).
\end{align}
To prepare the wave function at particular momentum point $\bk$, classically we just need to diagonalize the above Hamiltonian and occupy the two lowest eigenstates. To complete such a task in quantum computers, we need to find an efficient way to construct the corresponding quantum circuits. Even if the mathematics is straightforward, to generically construct the quantum circuits for a system with size larger than $3$ qubits, we resort to VQE methods \cite{Malley_PRX_2016, Kandala_Nature_2017,Kokail_Nature_2019,Yuan_quantum_2019,Grimsley_NC_2019,Lee_JCTC_2019,Chen_JCTC_2021,Yao_PRR_2021}. To use adaptive VQE to prepare the two-body ground state, we choose the initial state as the two-body ground state for the Hamiltonian given in Eq.~(\ref{Hmatrix}) with \emph{only the diagonal components}. By using the conservation of particle number of the model, the initial state can be easily obtained by applying two \textsf{X} gates onto the two qubits, which represents the two lower energy eigenstates. Then we restore the off-diagonal components of the Hamiltonian and implement the standard adaptive VQE steps to prepare the two-body ground state of the Hamiltonian specified by Eq.~(\ref{Hmatrix}). The corresponding circuit for these procedures is illustrated in Fig.~\ref{fig6}(a). The implementation details are discussed in the literature \cite{Grimsley_NC_2019} and can also be found in Appendix \ref{appendix_d1}. 

To evolve the wave function from $\bk$ to $\bk+\delta \bk$, we similarly begin with the approximate wave function at $\bk$ (obtained by the adaptive VQE described in the above) and then implement another adaptive VQE that minimizes the expectation values of the Hamiltonian density $H(\bk+\delta\bk)$. The resulted circuit obtained by these processes is shown in Fig.~\ref{fig6}(b). For both variational procedures, we used a convergence criterion of $\epsilon=0.01$.

A crucial observation is that due to the continuity of the wave function, the difference between the wave functions at two neighboring points is small, so the VQE corresponding to the wave function evolution can be completed with only a few operators from the pool (typically $ \lesssim 3$). Given the fact that the single excitation can be written as the sum of two Pauli strings $\mathcal{P} \in \{I,\sigma_x,\sigma_y,\sigma_z\}^{\otimes 3}$, the quantum circuit to measure $U_{\delta\bk}(\bk)$ can be further simplified by replacing the Hadamard test by direct projective measurements \cite{Mitarai_PRR_2019}, as shown in Fig.~\ref{fig6}(c) (see Appendix \ref{appendix_e} for details). Such replacements can reduce the depth of the circuit significantly at the cost of increasing the number of circuits by a factor of $\sim10$.

To control the circuit depth, and to have a measure of how far away the approximate wave functions are from the optimal ones, we introduce a threshold $\Delta E$, different from the convergence criterion $\epsilon$. We truncate the adaptive VQE circuit when the expectation value of the target Hamiltonian is within $\Delta E$ of the optimal one (see Appendix \ref{appendix_d2} for more details). It allows for a controllable balance of wave function accuracy and circuit depth. In Fig.~\ref{fig6}(d) and (e), we show the measured $n(\bk)$ on  IBMQ-Montreal within the magnetic BZ for $\Delta E = 0.2$ and $\Delta E =0.3$, respectively. By summing $n(\bk)$ over the whole magnetic BZ, we found that the Chern number of the two-body ground state is $-1$, which is consistent with the known result \cite{Fukui_JPSJ_2005}. We repeated the measurements another $4$ times on IBMQ-Montreal for both $\Delta E = 0.2$ and $\Delta E =0.3$, and obtained the same value(s). The detailed results for other measurements can be found in Appendix \ref{appendix_g}.

We also perform noisy simulations to determine how the measured results are affected by the machine noise. As described in the last section, we introduce the depolarizing errors to both single-qubit gates and two-qubit gates with $\varepsilon_2 = 10 \varepsilon_1$. The mistake ratio is plotted as a function of $\varepsilon_1$ and shown for $\Delta E=0.2$ and $\Delta E=0.3$ in Fig.~\ref{fig6}(f). For $\Delta E = 0.3$, we found that the error in measurements begins to appear only when $\varepsilon_1 \geq 0.009$, while for $\Delta E=0.2$, the error in measurements begins to appear at a smaller value $\varepsilon_1\geq0.007$. These results illustrate the crucial role of $\Delta E$ in achieving error-free measurements. If $\Delta E$ is too large, the true ground state is not well approximated. On the other hand, if $\Delta E$ is too small, the circuit depth becomes so large that the errors in the implementation of the circuit go beyond the tolerance of the topological protection, \textit{i.e.} the error in the measurement of $\tilde{F}$ is beyond $2\pi$. 

We need to emphasize that the parameter $\Delta E$ introduced here is to characterize the energy difference between the true ground state and the truncated state that we used to calculate Chern numbers for this particular model. It is equivalent to introduce a larger convergence criterion $\epsilon$, but in that case we will not know how far away the truncated states used for calculating Chern numbers are from the true ground states. One should view $\Delta E$ as similar to infidelity of the ground state, which is not readily available on a quantum computer, because we do not know what the true ground-state is, in general.

\subsection{Determining the topological phase diagram for the flux-$2\pi/3$ Quantum Hall model with a Hatsugai-Kohmoto interaction} \label{sec4b}

\begin{figure}[htpb]
  \centering
  \includegraphics[width=1\columnwidth]{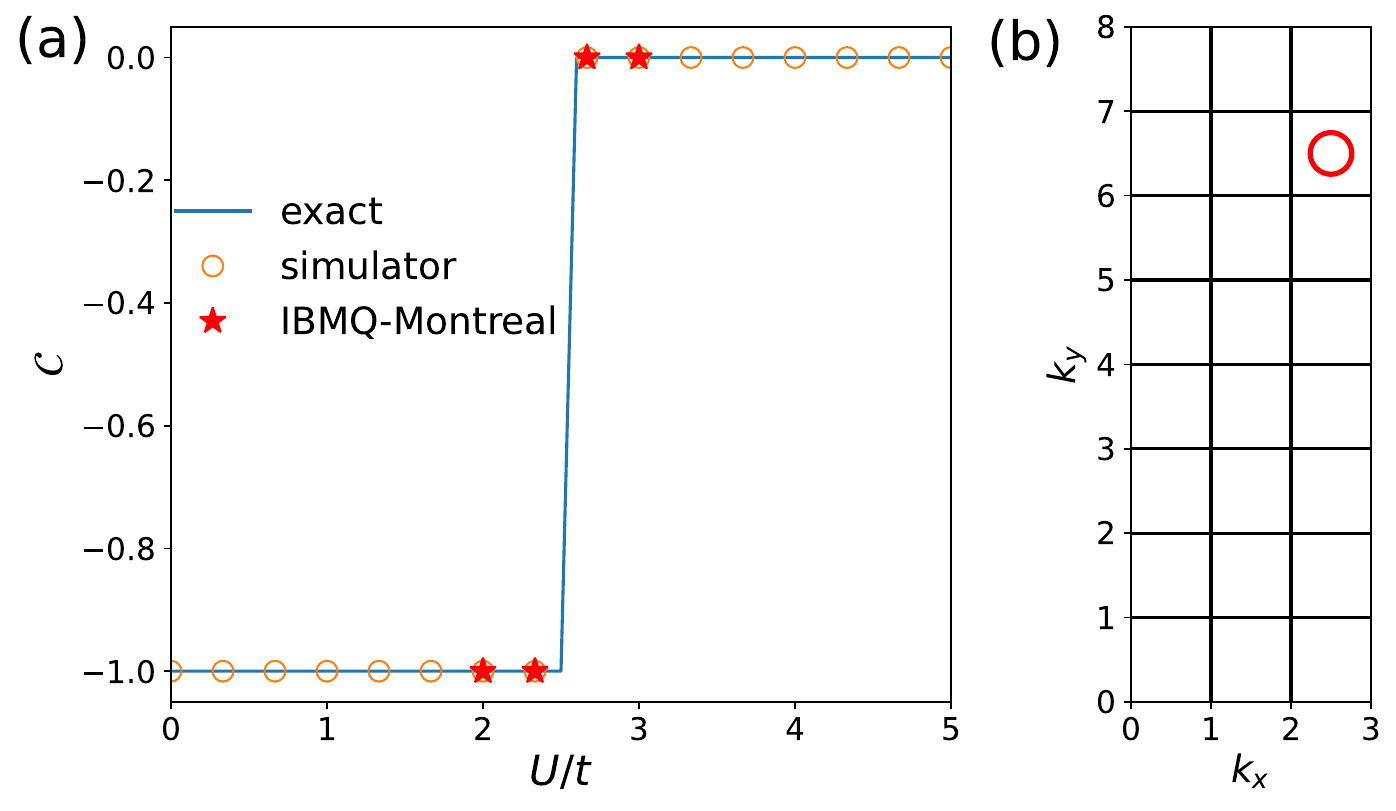}
  \caption{Calculations of the Chern number via adaptive VQE for the two-body ground state in the flux-$2\pi/3$ quantum Hall model with a Hatsugai-Kohmoto interaction: (a) the Chern number $\mathcal{C}$ of the model as a function of interaction strength $U$, from which a clear topological phase transition is induced by the Hatsugai-Kohmoto interaction between orbital $1$ and $2$. (b) The calculated integer-valued field $n(\bk)$ distribution in the discretized magnetic BZ obtained from IBMQ-Montreal for the $U=7/3$ case. In all the calculations, $N=8000$ were used to obtain the wave function overlaps.}
   \label{fig7}
\end{figure}

To demonstrate that our strategy works for interacting models, we introduce a Hatsugai-Kohmoto interaction to the flux-$2\pi/3$ Quantum Hall model. The Hatsugai-Kohmoto interaction is an interaction between two particles at the same momentum point \cite{Hatsugai_JPSJ_1992,Philips_NP_2020}, and here we introduce such an interaction for the particles at the orbitals $1$ and $2$:
\beq \label{HK_interaction}
H_{int} = U \sum_{\bk} n_{\bk;1} n_{\bk;2},
\eeq
where $n_{\bk;i} = c_{\bk;i}^\dag c_{\bk;i}$ with the orbital index $i=1,2$.

As we have discussed in the last section, the non-interacting limit of the model (the $U=0$ case), the two-body ground state of the model is topological with a Chern number $\mathcal{C}=-1$. The introduction of the Hatsugai-Kohmoto interaction results in a quantum phase transition from a topological phase with $\mathcal{C}=1$ to a trivial phase with $\mathcal{C}=0$, which can be seen from the plot of the Chern number as a function of the interaction strength $U$ shown in Fig.~\ref{fig7}(a).

Similar to the last subsection, we adopt adaptive VQE to calculate the Chern number on quantum hardware. First we notice that the model has three orbitals, so according to the unitary coupled cluster ansatz double-excitation operators do not exist (for fermionic systems, double-excitation operators appear when the system has four or more qubits). Therefore, the operator pool for the present calculations is the same with the last section, and the only difference is that in the optimization we need to add the new interaction term specified by Eq.~(\ref{HK_interaction}).

We continue to use projective measurements to replace the evolution circuit as shown in Fig.~\ref{fig6}(c), which can efficiently reduce the circuit depth for small systems. In the present optimization, we simply introduce a larger convergence criterion $\epsilon=0.1$ and do not characterize how far away the prepared state is from the true ground state. With the concrete quantum circuits, we use them to calculate the Chern number for this interaction model. The results from quantum simulators were shown as the orange open circles in Fig.~\ref{fig7}(a), which perfectly coincide with the exact values. At around the transition point, we performed the calculation of the Chern number with real quantum hardware IBMQ-Montreal; the results are shown as red stars in Fig.~\ref{fig7}(a). Again, the real machine data perfectly falls onto the exact values without doing any error corrections. 

The correct calculations of the Chern numbers on real hardware for this interacting model demonstrate that our holonomy strategy, combined with VQE, allows us to correctly determine the topological invariant for generic quantum states. The typical depth of quantum circuits for the calculations of Chern numbers for the flux-$2\pi/3$ Quantum Hall model (both with and without interactions) is $\sim 25$ and the number of \textsf{CNOT} gates is $\sim 30$. It is remarkable that quantum circuits with this depth obtain an exact measurement for any system quantity, and this is entirely due to the robustness
of wavefunction topology to mild deformations.


\subsection{Determining the topological phase diagram for interacting Chern insulators} \label{sec4c}

The integration of our strategy for measuring topological invariants via holonomy with VQE enables a move towards more complex interacting models. This combination provides an immediate tool to prepare and characterize interacting topological states on quantum computers. In this subsection, we will demonstrate how to calculate the topological invariants for an interacting model; we will map out the quantum phase diagram of the interacting Chern insulators by using our holonomy strategy with VQE. 

We consider a model constructed on a square lattice with each unit cell containing two sublattices. We denote the two sublattices as the $A$ and $B$ sites. Then the Hamiltonian for a non-interacting Chern insulator can be written as \cite{Qi_PRB_2006}:
\begin{align}
H_{CI} &= -t\sum_{i} \left(\psi_{i}^\dag \frac{\sigma_z-i\sigma_x}{2} \psi_{i+\hat{x}} +  \psi_{i}^\dag \frac{\sigma_z-i\sigma_y}{2} \psi_{i+\hat{y}}\right) \nonumber \\
&-t\sum_{i} \left(\psi_{i}^\dag \frac{\sigma_z+i\sigma_x}{2} \psi_{i-\hat{x}} +  \psi_{i}^\dag \frac{\sigma_z+i\sigma_y}{2} \psi_{i-\hat{y}}\right) \nonumber \\
&+M \sum_{i} \psi_{i}^\dag \sigma_z \psi_{i},
\end{align}
where $\psi_i = \left( c_{i,A}, c_{i,B}\right)^T$ with $c_{i,A}$ and $c_{i,B}$ the annihilation operators for the sites $A$ and $B$ in the unit cell $i$. Here we assume that the interaction happens for the particles within a unit cell:
\beq
H_{int} = U \sum_{i} n_{i,A} n_{i,B},
\eeq
where $n_{i,\alpha} = c_{i,\alpha}^\dag c_{i,\alpha}$ is the occupation number operator for the site $\alpha$ in the unit cell $i$.

\begin{figure}[t]
  \centering
  \includegraphics[width=1\columnwidth]{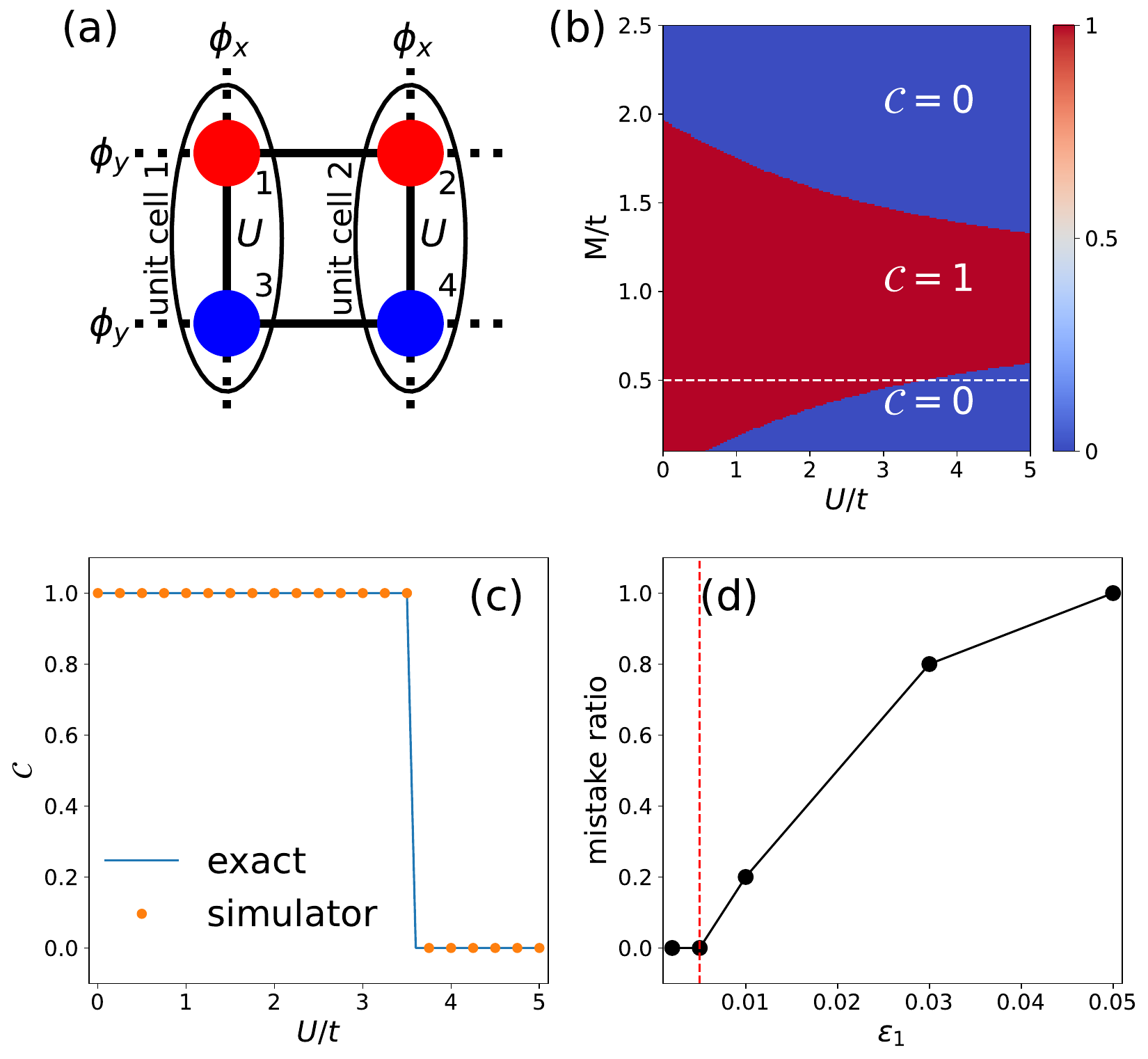}
  \caption{Calculations of the Chern number via adaptive VQE for the interacting Chern insulator model: (a) the lattice configuration of the $4$-site interacting Chern insulator model (containing two unit cells) with the interaction U applied to the fermions within the same unit cell; (b) the interacting quantum phase diagram of the model on the $M$-$U$ parameter plane determined by the exact diagonalization method; (c) the calculated Chern number along the white dashed line in (b), on which the results obtained from quantum simulator are consistent with the exact diagonalization results; (d) the mistake ratio as a function of the depolarizing single-qubit error rate $\epsilon_1$ determined by $n=10$ noisy simulations with the two-qubit error rate $\epsilon_2=10\epsilon_1$. For the calculations on quantum simulator and noisy simulator, $N=3072$ were used to obtain the wave function overlaps. The convergence parameter is $\varepsilon=0.002$ because of the small excitation energy gap in this model.}
   \label{fig8}
\end{figure}

In the following explicit example, we studied the model constructed on a finite-size lattice with two unit cells denoted by $i=1$ and $i=2$ [see Fig.~\ref{fig8}(a)]. To calculate Chern numbers of the interacting model, we need to introduce twisted boundary conditions for the $x$ and $y$ directions, which can be imposed by introducing hopping phases across the boundaries of the finite-size lattice \cite{Niu_prb_1985}; we denote these by $\phi_x$ and $\phi_y$ in Fig.~\ref{fig8}(a). With the twist boundary conditions, the Hamiltonian describing the system can be explicitly written into the sum of two components:
\beq \label{interacting_Chern_Hamiltonian}
H_{tot}(\phi_x,\phi_y) = H_{bulk} + H_{bdy}(\phi_x,\phi_y),
\eeq
where the phase factors $\phi_x$ and $\phi_y$ denotes the twist angle specifying the boundary conditions, and $H_{bulk}$ and $H_{bdy}$ are given explicitly as:
\begin{align}
H_{bulk} = &-t \left[ \psi_{1}^\dag \left( \sigma_z - i\sigma_x \right) \psi_{2} + h.c. \right] + M \sum_{i=1,2} \psi_{i}^\dag \sigma_z \psi_{i} \nonumber \\
& +U \sum_{i} n_{i,A} n_{i,B},
\end{align}
and
\begin{align}
H_{bdy} = &-\frac{t}{2} \left[e^{-i\phi_x} \psi_{1}^\dag \left( \sigma_z - i\sigma_x \right) \psi_{2} + h.c. \right] \nonumber \\
&-\frac{t}{2} \left[ e^{-i\phi_y}  \psi_{1}^\dag  \left( \sigma_z - i\sigma_y \right) \psi_{1} + h.c. \right] \nonumber \\
&-\frac{t}{2} \left[ e^{-i\phi_y}  \psi_{2}^\dag  \left( \sigma_z - i\sigma_y \right) \psi_2 + h.c. \right].
\end{align}
The parameter space made by the twist angles $\boldsymbol{\phi} = (\phi_x, \phi_y)$ plays the same role as a Brillouin zone, and to use the holonomy strategy we need to discretize the space into a $N_x \times N_y$ mesh, with the distance between neighboring points denoted as $\delta \phi_x$ in the $x$ direction or $\delta \phi_y$ in the $y$ direction. Following the holonomy strategy, once we obtained the wave functions at two neighboring points in the parameter space, we can similarly define the normalized overlap $U_{\delta \boldsymbol{\phi}}(\boldsymbol{\phi}) = \langle \Psi(\boldsymbol{\phi}) | \Psi(\boldsymbol{\phi}+\delta \boldsymbol{\phi}) \rangle / |\langle \Psi(\boldsymbol{\phi}) | \Psi(\boldsymbol{\phi}+\delta \boldsymbol{\phi}) \rangle|$ with $\delta \boldsymbol{\phi} = \delta \phi_x \hat{x}$ or $\delta \boldsymbol{\phi} = \delta \phi_y \hat{y}$. Then the Berry curvature in the twist-angle space can be defined similarly as:
\begin{align}
\mathcal{F}(\boldsymbol{\phi}) = \ln \left[ \frac{U_{\delta \hat{\boldsymbol{\phi}}_x} (\boldsymbol{\phi}) U_{\delta \hat{\boldsymbol{\phi}}_y}(\boldsymbol{\phi}+\delta \hat{\boldsymbol{\phi}}_x )}{U_{\delta \hat{\boldsymbol{\phi}}_x}(\boldsymbol{\phi}+\delta \hat{\boldsymbol{\phi}}_y ) U_{\delta \hat{\boldsymbol{\phi}}_y}(\boldsymbol{\phi})}  
\right].
\end{align}
With the Berry curvature in hand, the Chern number can be calculated by summing over the Berry curvature at each discretized mesh point in the twist-angle space, similar to Eq.~(\ref{Eq1}). 
Fig.~\ref{fig8}(b) shows the phase diagram obtained
from classical calculation,
where we observe a narrowing of the topological
region as a function of interaction strength.


To prepare the ground state wave functions at each $\boldsymbol{\phi}$, we use the adaptive VQE algorithm based on the unitary coupled cluster theory.
However, different from the example of quantum Hall states shown in Sec. \ref{sec4a}, to account for
the effects of the interaction double excitation operators are necessary. We thus truncate the operator
pool at this level (for the details of the operator pool for this model see Appendix \ref{appendix_f}) \cite{Lee_JCTC_2019}. As in the above, each normalized wave function overlap can be calculated by using the general circuit shown in Fig.~\ref{fig1}(c) with the wave function preparation part and the wave function evolution part realized by adaptive VQE algorithm and projective measurements as shown in Fig.~\ref{fig6}(c).

In Fig.~\ref{fig8}(c), we compare the results obtained from quantum simulators and those from exact diagonalization along the white dashed line (with $M=0.5$) in Fig.~\ref{fig8}(b). The consistency of the two results indicates that our strategy correctly calculates Chern numbers even for interacting models on quantum computers. 

Finally, we performed noisy simulations to determine the error-free threshold for the calculation of the Chern numbers of this model. We still set the depolarizing single-qubit error $\epsilon_1$ as $10$ times of the two-qubit error $\epsilon_2$, and for each $\epsilon_1$ $10$ trials were performed. The noisy simulations determine the error-free threshold for the Chern number calculations for this model is $\epsilon_1=0.005$, as indicated by the red dashed line in Fig.~\ref{fig8}(d). The decrease of the error-free threshold is due to two reasons. First, this model has a very small excitation energy gap and requires more iteration steps to capture the necessary features of ground state. Second the double-excitation operators are needed to achieve good convergence, and as we show in Appendix \ref{appendix_f}, the circuit depth for one double-excitation operator would is $40$ CNOTs, which is much longer than that of a single-excitation operator.

\section{Discussion and Remarks} \label{sec5}

We end by comparing our method with the recently proposed scheme \cite{Cian_arXiv_2020} to measure the many-body Chern number by randomized measurements. In Ref.~\cite{Cian_arXiv_2020}, the many-body Chern number is inferred from the winding of the measured expectation of the SWAP operator applied to the two copies of the wave function after `surgery', while our method can measure the Chern number directly avoiding the difficulties in inferring winding (see the discussion of measuring the winding of Zak phase). Though similar to \cite{Cian_arXiv_2020}, the demonstrations of our strategy for interacting models are performed on quantum hardware or quantum simulator, and, more importantly, the interacting model studied in our work has explicitly included interaction terms, which, however, is implicit as the hardcore assumptions in \cite{Cian_arXiv_2020}.

In our work, we have explicitly demonstrated that our strategy allows the calculations of topological invariants in general as long as the state can be prepared by an UCC-based adaptive VQE algorithm \cite{Grimsley_NC_2019}. In fact our strategy can be also easily incorporated with other VQE algorithms, {\it i.e.} the qubit coupled cluster version \cite{Ryabinkin_JCTC_2018} or VQE algorithms based on subspace expansion \cite{Motta_NP_2020}. In this sense, whether our strategy can perform calculations beyond the ability of classical computers depends on the further development of VQE algorithms. Unfortunately, at the present stage, the VQE algorithms still need a large number of multi-qubit gates and fails on real hardware for large-size problems, unless the error rates on quantum hardware can be significantly improved \cite{Fedorov_arXiv_2021}. How to develop an efficient VQE algorithm, which could perform calculations beyond the ability of classical computers, is an active research field and beyond the focus of the present work, but the present work demonstrate a prototype that our strategy can be incorporated with more advanced VQE algorithms to solve problems beyond classical computers in the future.

In conclusion, by showing the error-free calculations of Chern numbers for both non-interacting models and interacting models on IBM machines, we demonstrated how the topological properties of wave functions can be robustly measured by quantum computers in this NISQ era. The proposed strategy and its integration with VQE can provide a powerful tool to investigate various topologically ordered systems on current quantum hardware.

\section*{Acknowledgment}

We acknowledge helpful discussions with Michael Geller. We acknowledge use of the IBM Q for this work. The views expressed are those of the authors and do not reflect the official policy or position of IBM or the IBM Q team. Access to the IBM Q Network was obtained through the IBM Q Hub at NC State. We acknowledge the use of the QISKIT software package~\cite{qiskit2019} for performing the quantum simulations. This work was supported by the Department of Energy, Office of Basic Energy Sciences, Division of Materials Sciences and Engineering under Grant No. DE-SC0019469. J.K.F. was also supported by the McDevitt bequest at Georgetown. The data for the figures and sample codes are available at \url{https://doi.org/10.5061/dryad.xsj3tx9g5}.

\appendix

\section{Noise model simulations} \label{appendix_a}

In our noise simulations, we adopt a minimal noise model incorporating only the depolarizing noise for the single-qubit and two-qubit gates, which is a widely used method to incorporate quantum errors \cite{Knill_Nature_2005,Cross_QIC_2009}. This quantum depolarizing noise for the $n$-qubit gate operation can be written as:
\beq
\Delta_{\lambda}(\rho) = \left( 1-\lambda \right) \rho + \lambda \mathrm{Tr}\left[ \rho \right] \frac{I}{2^{n}},
\eeq 
where $\lambda$ is the depolarization error parameter, $n$ is the number of qubits, and $\rho$ can be regarded as the density matrix corresponding to the operation. This depolarizing noise model has been included in the noise model module of qiskit \cite{qiskit2019}, and we used this module directly for our noise simulations.

However, in real quantum hardware there are still other kinds of error, such as the readout error and thermal relaxation error. By the direct comparison with the calibration data shown in Table \ref{tab:table-calibration_toronto1}, \ref{tab:table-calibration_toronto2}, and \ref{tab:table-calibration_montreal}, we can observe that the parameters we used in noise simulations are much larger than the calibration data of gate errors. In this sense, our noise model accounts for errors beyond those in the calibration, and thus is an effective model to test how the machine noise affects the measurement of topological invariants.

\section{The true error-free threshold for the calculation of Chern numbers in the chiral $p$-wave superconducting model} \label{appendix_b}

\begin{figure}[!t]
  \centering
  \includegraphics[width=1\columnwidth]{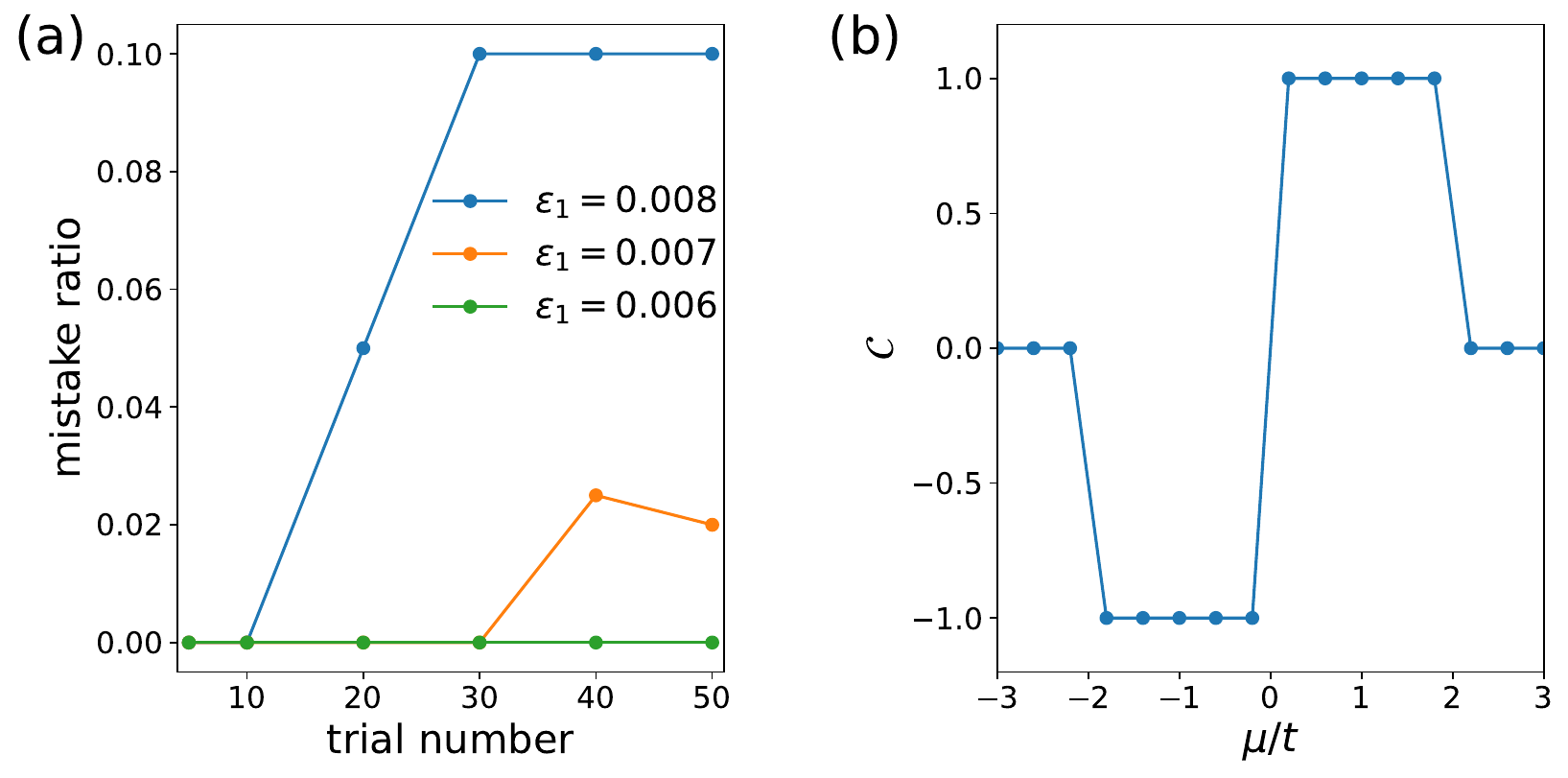}
  \caption{Determining the true error-free threshold: (a) the mistake ratio as a function of trial numbers for three different error rates; (b) the average of $100$ trials of noisy simulation results for $\epsilon_1=0.006$. In the calculations, we still assume the single-qubit error rate $\epsilon_1$ is $10$ times of two-qubit error rate $\epsilon_2$ so that the calculations are controlled by the single-qubit error rates.}
   \label{figs1}
\end{figure}

In Fig.~\ref{fig2} of the main text, we have performed noisy simulations to determine the error-free threshold in the presence of depolarizing errors. The data presented in that figure was obtained by $10$ trials, which might be not large enough to determine the true error-free threshold. In this appendix, we further determine the true error-free threshold by increasing the trial numbers. The simulation results were summarized in Fig.~\ref{figs1}. We found that the error-free threshold decreases and converges to $\epsilon_1=0.006$ (see Fig.~\ref{figs1}). For the case with $\epsilon_1=0.006$, we found that no error happens even with $100$ trials, which is shown in Fig.~\ref{figs1}.

\section{The noisy simulations close to critical points} \label{appendix_b1}

To examine the influence of shot noise, we performed noisy simulations for the systems close to the topological critical points of the chiral $p$-wave superconducting model, where the admissibility is much easier to break. 

\begin{figure}[h]
  \centering
  \includegraphics[width=1\columnwidth]{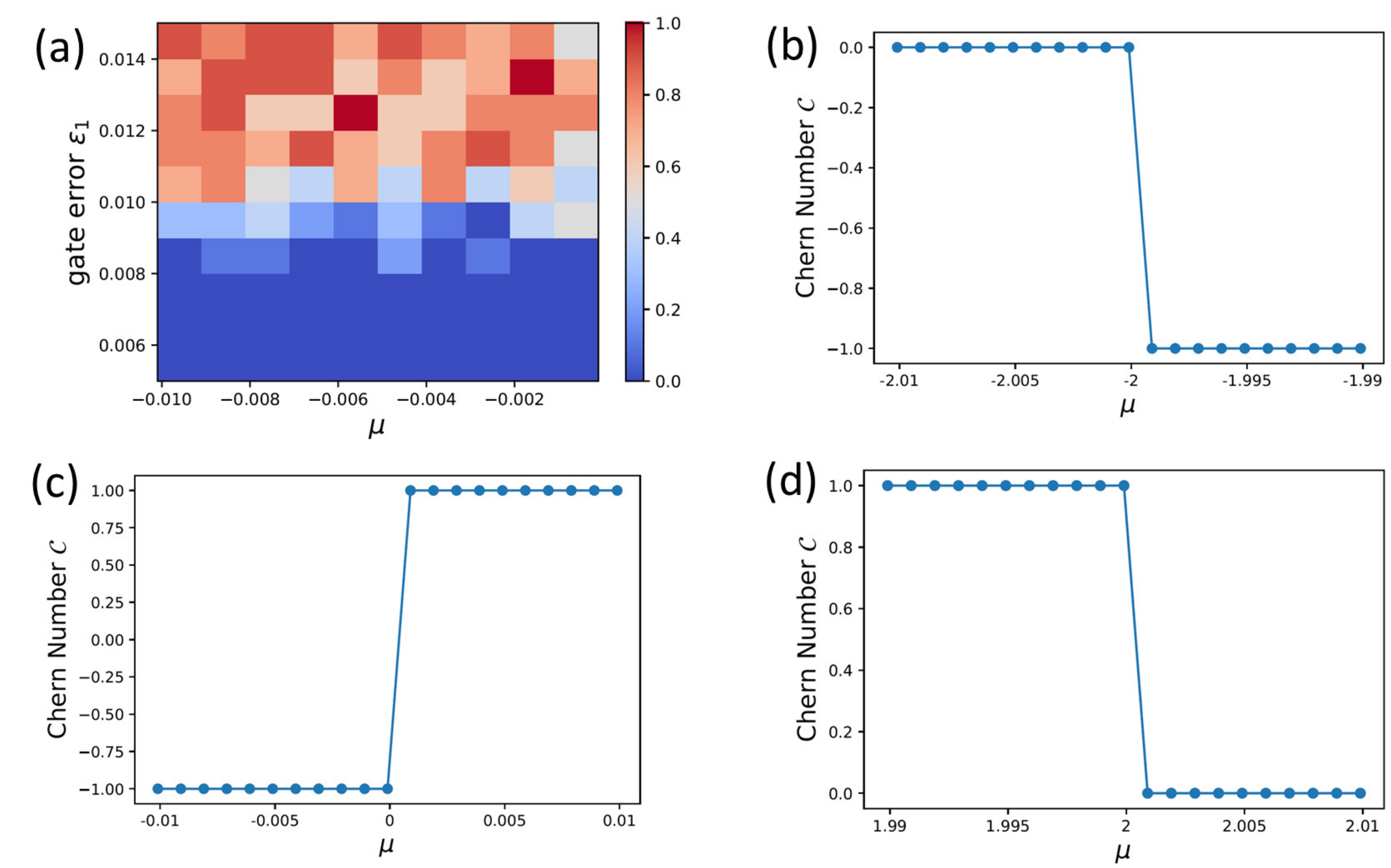}
  \caption{Noise simulations of the calculations of Chern numbers at around critical points: (a) the mistake ratios of Chern number measurement from the noise simulations. The calculations were performed at 10 evenly distributed points from $\mu=-0.0101$ to $\mu=-0.0001$. For each error rates and $\mu$, 10 calculations were performed. (b) the mean values of Chern numbers calculated at around the critical point $\mu_c=-2$ with $50$ repeated calculations for $\epsilon_1=0.006$. The calculations were performed from $\mu=-2.0101$ to $\mu=-1.9901$ with a step $0.001$. (c) the mean values of Chern numbers calculated at around the critical point $\mu_c=0$ with $50$ repeated calculations for $\epsilon_1=0.006$. The calculations were performed from $\mu=-0.0101$ to $\mu=0.0099$ with a step $0.001$; (d) the mean values of Chern numbers calculated at around the critical point $\mu_c=2$ with $50$ repeated calculations for $\epsilon_1=0.006$. The calculations were performed from $\mu=1.9901$ to $\mu=2.0101$ with a step $0.001$.}
   \label{figsn}
\end{figure}

The same with the assumptions used in the main text, we assumed that the two-qubit error $\epsilon_2$ is 10 times of the single-qubit error $\epsilon_1$. We first determined the error-free threshold, when the systems are close to the critical point $\mu_c=0$. The calculations were performed by verifying the single-qubit error from $\epsilon_1=0.005$ to  $\epsilon_1=0.015$ with a step $0.001$. The parameter controlling the system $\mu$, was tuned from $\mu=-0.0101$ to $\mu=-0.0001$ with a step length $0.001$. From Fig. \ref{figsn}(a) we found that the error-free threshold does not change, when the system get closer to the critical points.

From the discussion in Appendix \ref{appendix_b}, we know that with the increasing number of repeated noisy simulation, the error-free threshold reduces and seems to stabilize at $\epsilon_1=0.006$, when the number of repeated calculations is larger than $50$. To further verify the possible influence of shot noise, we set $\epsilon_1=0.006$ and repeat the calculations $50$ times for each $\mu$ close to all the three critical points of the model $\mu_c=\{-2,0,2\}$. The results were shown in Fig. \ref{figsn}(b), (c) and (d). We found that even when $|\mu-\mu_c|=0.0001$, no error can be found (since the average of 50 calculations gives the correct Chern number without any deviations).

\section{Use of the extra Toffoli gate in the interband evolution circuit in Fig.~2(b)} \label{appendix_c}

The interband evolution appears in the calculations of the topological invariant for mixed states, {\it i.e.} the ensemble geometric phase discussed in the main text. Explicitly, when we use Eq.~(\ref{EGP}) to calculate the ensemble geometric phase, the central quantity is $M_T$ defined in Eq.~(\ref{def_MT}). According to Eq.~(\ref{def_MT}), the wave function overlap is now given by the inner product of the wave functions for all the energy bands in the Bloch basis (see \cite{Bardyn_PRX_2018} for more details), which can be written explicitly in the matrix form as:
\begin{widetext}
\beq
\mathcal{U}_{\bk+\delta k_x \hat{x}}^\dag  \mathcal{U}_{\bk} = \left(\begin{array}{cccc} 
\Psi_1^\dag(\bk+\delta k_x \hat{x}) \Psi_1(\bk) & \Psi_1^\dag(\bk+\delta k_x \hat{x}) \Psi_2(\bk) & \cdots & \Psi_1^\dag(\bk+\delta k_x \hat{x}) \Psi_{\mathcal{N}}(\bk) \\
\Psi_2^\dag(\bk+\delta k_x \hat{x}) \Psi_1(\bk) & \Psi_2^\dag(\bk+\delta k_x \hat{x}) \Psi_2(\bk) & \cdots & \Psi_2^\dag(\bk+\delta k_x \hat{x}) \Psi_{\mathcal{N}}(\bk) \\
\vdots & \vdots & \ddots & \vdots \\
\Psi_{\mathcal{N}}^\dag(\bk+\delta k_x \hat{x}) \Psi_1(\bk) & \Psi_{\mathcal{N}}^\dag(\bk+\delta k_x \hat{x}) \Psi_2(\bk) & \cdots & \Psi_{\mathcal{N}}^\dag(\bk+\delta k_x \hat{x}) \Psi_{\mathcal{N}}(\bk)
 \end{array}\right),
\eeq 
\end{widetext}
where the diagonal elements of the matrix are implemented by the intraband evolution operations, while the off-diagonal elements indicate the evolution between different bands and will be implemented by the interband evolution operations.

To find the interband evolution, we begin with the wave function at the momentum point $\bk$ for one of the bands $\Psi_{\alpha}(\bk)$, and the interband evolution circuit is to transform the state to $\Psi_{\bar{\alpha}}(\bk+\delta \bk)$. Recall that the present model is a two-band model, and $\bar{\alpha}=+/-$ when $\alpha=-/+$. To realize the transformation, we first evolve the wave function back the to band space by implementing $V_\alpha^\dag(\bk)$ (see the discussion around Eq.~(26)). In the band space, the lower band wave function is simply $\left(\begin{array}{cc} 0 & 1\end{array}\right)^T$ and the upper band wave function is $\left(\begin{array}{cc} 1 & 0\end{array}\right)^T$. Therefore, to transform between the two bands, we need to implement a \textsf{SWAP} gate. Finally we just need to implement $V_{\bar{\alpha}}(\bk+\delta \bk)$ to obtained the desired state.

Using the decomposition of the \textsf{SWAP} gate into \textsf{CNOT} gates,
\begin{align}
\textsf{SWAP}[q_0,q_1] = \textsf{CNOT}[q_1,q_0] \textsf{CNOT}[q_0,q_1] \textsf{CNOT}[q_1,q_0],
\end{align}
we find:
\begin{align}
&V_\alpha^\dag(\bk) \textsf{SWAP}[q_0,q_1] V_{\bar{\alpha}}(\bk+\delta \bk) \nonumber \\
= & \textsf{CNOT}[q_1,q_0] \mathrm{CU}_3[q_0,q_1](\theta_{-\bk},-\varphi_{-\bk},\varphi_{-\bk}) \textsf{CNOT}[q_0,q_1] \nonumber \\
\times &  \mathrm{CU}_3[q_0,q_1](\theta_{\bk+\delta \bk},-\varphi_{\bk+\delta \bk},\varphi_{\bk+\delta \bk}) \textsf{CNOT}[q_1,q_0].
\end{align}
To implement the Hadamard test, we just need to use an ancilla to control all the operations given in the above equations. The expression can be represented by the quantum circuit shown in Fig.~\ref{fig2}(b).

\begin{widetext}

\section{The implementation of adaptive VQE for the quantum Hall model} \label{appendix_d}

To motivate the variational ansatz, we can further write the Hamiltonian as:
\begin{align} \label{Eq_tb}
H &= \sum_{\bk} \left[ -2 \cos k_y c_{\bk;1}^\dag c_{\bk;1} -2 \cos(k_y+2\pi/3) c_{\bk;2}^\dag c_{\bk;2} -2 \cos(k_y+4\pi/3) c_{\bk;3}^\dag c_{\bk;3}\right] \nonumber \\
& - \sum_{\bk} \left[\left(c_{\bk;1}^\dag c_{\bk;2} + c_{\bk;2}^\dag c_{\bk;1}\right) +  \left(c_{\bk;2}^\dag c_{\bk;3} + c_{\bk;3}^\dag c_{\bk;2}\right) + \left(e^{-i3k_x}c_{\bk;1}^\dag c_{\bk;3} + e^{i3k_x}c_{\bk;3}^\dag c_{\bk;1}\right)\right],
\end{align}
\end{widetext}
where the first line on the right hand side contains the `on-site' terms determining the Hartree-Fock ground state for $\bk$, and the second line contains all possible terms generating single excitations from the Hartree-Fock ground state. Here the magnetic BZ is defined with $k_x \in [0,2\pi/3]$ and $k_y \in [0,2\pi]$. Of note is the last term in the second line, which is different from the conventional hopping. It can be regarded as a `spin-orbit' fashion and gives rising to the non-trivial topology of the model. 

For this model, there are three bands, and a straightforward calculation shows that the Chern numbers associated with them are $1$, $-2$, $1$ from the low energy to high energy bands. Therefore, one can easily identify that the Chern number of the two-body ground state of this model has a Chern number $1-2=-1$. Because we have set the hopping strength as the energy unit, the exciation gap above the two-body ground state is $\sim 2$ for this model.

\subsection{The procedures for implementing the adaptive VQE in the quantum Hall model} \label{appendix_d1}

Adaptive VQE is a hybrid quantum-classical algorithm; the evaluation of expectation values are carried out on quantum hardware, and the optimization is performed on classical computers. 

\begin{figure*}[!t]
  \centering
  \includegraphics[width=1.6\columnwidth]{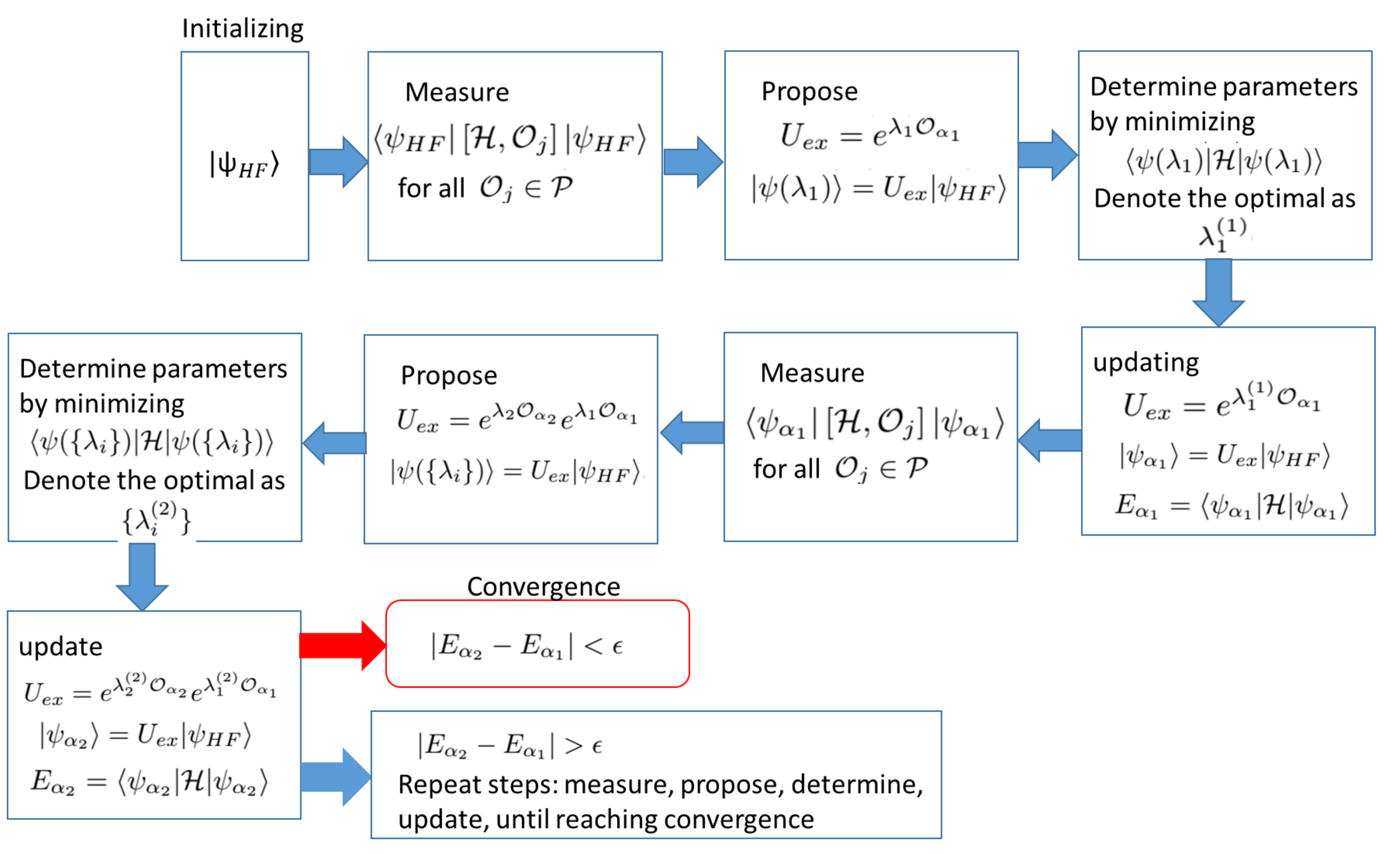}
  \caption{The flow diagram for the implementation of adaptive VQE.}
   \label{figs2}
\end{figure*}

\subsubsection{Determining the operator pool in the adaptive VQE}

This particular model contains only hopping terms, so it can be straightforwardly solved by generalizing the factorized form of the unitary coupled cluster (UCC) theory (truncated at single excitations). We choose the operator pool of the adaptive VQE to include all the possible operations generating single excitations/de-excitations, constrained by the Hamiltonian. In fermionic language, these operators are:
\beq
\begin{cases}
\mathcal{O}_1 = c_{\bk;1}^\dag c_{\bk;2} - c_{\bk;2}^\dag c_{\bk;1}, \\
\mathcal{O}_2 = c_{\bk;2}^\dag c_{\bk;3} - c_{\bk;3}^\dag c_{\bk;2}, \\
\mathcal{O}_3 = c_{\bk;3}^\dag c_{\bk;1} - c_{\bk;1}^\dag c_{\bk;3}, \\
\mathcal{O}_4 = i\left(c_{\bk;3}^\dag c_{\bk;1} + c_{\bk;1}^\dag c_{\bk;3}\right),
\end{cases}
\eeq
where $c_{\bk;n}$ with $n=1,2,3$ denotes the annihilation operator for the state in the orbital $n$ with momentum $\bk$.
So, the operator pool is given by $\mathcal{P} = \{ \mathcal{O}_1, \mathcal{O}_2, \mathcal{O}_3, \mathcal{O}_4 \}$. The Jordan-Wigner transformation can be used to map them to qubit representations:
\beq \label{JW}
\begin{cases}
\mathcal{O}_1 = i\left(\sigma_3^0 \otimes \sigma_2^x \otimes \sigma_1^x + \sigma_3^0 \otimes \sigma_2^y \otimes \sigma_1^y\right)/2, \\
\mathcal{O}_2 = i\left(\sigma_3^x \otimes \sigma_2^x \otimes \sigma_1^0 + \sigma_3^y \otimes \sigma_2^y \otimes \sigma_1^0\right)/2, \\
\mathcal{O}_3 = i\left(\sigma_3^x \otimes \sigma_2^z \otimes \sigma_1^x + \sigma_3^y \otimes \sigma_2^z \otimes \sigma_1^y\right)/2, \\
\mathcal{O}_4 = -\left(\sigma_3^x \otimes \sigma_2^z \otimes \sigma_1^y - \sigma_3^y \otimes \sigma_2^z \otimes \sigma_1^x\right)/2,
\end{cases}
\eeq
where $\sigma^\alpha_k$ indicates the $\alpha-$th Pauli matrix acting on the $k$-th orbital ($\alpha=0$ is the identity).

Note that the choice of the operator pool is different from the conventional UCC theory in the following ways: 1. in the conventional UCC theory, the single-excitations are generated by the hopping between an occupied orbital and an empty orbital, but here we include all the hopping operations; and 2. $\mathcal{O}_4$ is not included in the conventional UCC theory. The reasons for these differences are the following: first the occupied orbitals changes when we sweep the whole magnetic BZ, so a universal operator pool should include all the hopping terms, and the issue of the enlargement of the pool at each momentum point can be naturally solved by the adaptive VQE, which always chooses the most efficient operators in the pool to prepare the ground state wave function; secondly, the non-trivial topology of the model is due to the `spin-orbit' like hopping at the boundary of the magnetic BZ, so introducing $\mathcal{O}_4$ naturally accounts for this term.

With these components, an arbitrary wave function of the model can be generated by the following ansatz \cite{Grimsley_NC_2019}:
\beq
|\Psi_{target}\rangle = \prod_{\alpha} e^{\lambda_\alpha \mathcal{O}_\alpha} |\Psi_{initial}\rangle,
\eeq
where $|\Psi_{initial}\rangle$ denotes an initial wave function, $|\Psi_{target}\rangle$ is the target wave function, $\mathcal{O}_\alpha \in \mathcal{P}$ can be repeatedly appear in the ansatz, and $\lambda_\alpha$ is a real parameter to be determined in the optimization.

\subsubsection{Choosing a proper operator from the pool}

In the adaptive VQE \cite{Grimsley_NC_2019}, the key step is to find the best operator from the pool, to optimize the present wave function. This task can be done by the measurements of the expectation values of the commutators between the Hamiltonian density $\mathcal{H}$ and operators in the pool $\mathcal{O}_{\alpha}\in\mathcal{P}$. Suppose that the present wave function is given by $|\psi\rangle$. Then the expectations of the commutators are
\beq \label{gradient}
\frac{\partial E}{\partial \lambda_\alpha} = \langle \psi | \left[ \mathcal{H}, \mathcal{O}_\alpha \right] |\psi\rangle.
\eeq
Suppose that $|\langle \psi | \left[ \mathcal{H}, \mathcal{O}_{\alpha_0} \right] |\psi\rangle|$ is the largest of all the commutators. Then it indicates that $\mathcal{O}_{\alpha_0}$ currently leads to the steepest decent from the present expectation energy $E=\langle\psi|\mathcal{H}|\psi\rangle$. Therefore, $\mathcal{O}_{\alpha_0}$ is the best operator to be used in the optimization.

\subsubsection{Adaptive VQE procedure}

The aim of the adaptive VQE is to find an approximate operator, which can transform the initial Hartree-Fock ground state wave function determined by the first line of Eq.~(\ref{Eq_tb}), to the approximate ground-state wave function. If we suppose that the approximate operator can be found after $N$ iterations of adaptive VQE, it means that the exact operator preparing the true ground state wave function can be approximated by:
\beq
U_{ex} \approx e^{\lambda_N \mathcal{O}_{\alpha_N}} e^{\lambda_{N-1} \mathcal{O}_{\alpha_{N-1}}} \cdots e^{\lambda_1 \mathcal{O}_{\alpha_1}},  
\eeq
where $\lambda_{n}$ with $n\in\{1,2,\cdots,N\}$ is a real parameter, and $\mathcal{O}_{\alpha_n}$ is an operator in the operator pool $\mathcal{P} = \{\mathcal{O}_1,\mathcal{O}_2,\mathcal{O}_3,\mathcal{O}_4\}$. 
Note that operators are allowed to repeat.

\begin{figure}[h]
  \centering
  \includegraphics[width=0.9\columnwidth]{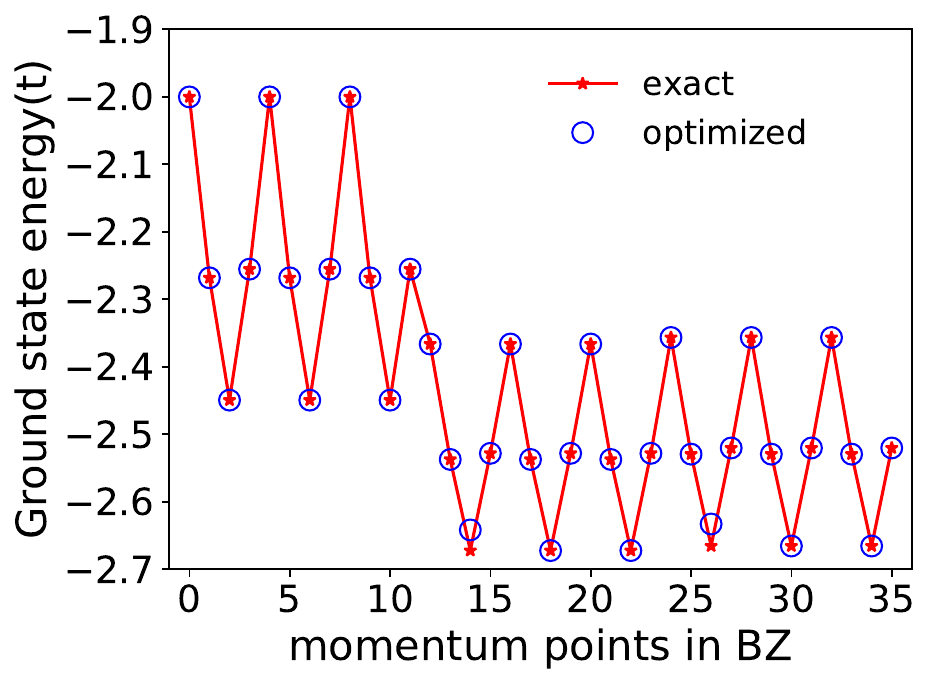}
  \caption{The comparison between the exact two-body ground state energy and the optimized energy by adaptive VQE. In the calculation, the criteria to stop the adaptive VQE is the same with that used in main text, which is $\epsilon=0.01$. The difference between the exact ground state energy and the optimized energy is smaller than 0.03. Here the energy unit is the hopping strength of the model.}
   \label{fig12}
\end{figure}

With the determination of the operator pool and the method to determine the best operators from the pool, we can implement the adaptive VQE explicitly, as summarized in Fig.~\ref{figs2} and illustrated in ref.\cite{Grimsley_NC_2019}:

\begin{enumerate}
    \item We begin with the Hartree-Fock ground state determined by the first line in Eq.~(\ref{Eq_tb}), which is denoted by $|\psi_{HF}\rangle$, and then we calculate: $\langle \psi_{HF} | \left[ \mathcal{H}, \mathcal{O}_1 \right] |\psi_{HF}\rangle$, $\langle \psi_{HF} | \left[ \mathcal{H}, \mathcal{O}_2 \right] |\psi_{HF}\rangle$, $\langle \psi_{HF} | \left[ \mathcal{H}, \mathcal{O}_3 \right] |\psi_{HF}\rangle$, and $\langle \psi_{HF} | \left[ \mathcal{H}, \mathcal{O}_4 \right] |\psi_{HF}\rangle$ to determine the maximum gradient (see Eq. (\ref{gradient})).

    \item Based on the expectation values evaluated in step 1, we can find the operator $\mathcal{O}_{\alpha_1} \in \mathcal{P}$, which makes $|\langle \psi_{HF} | \left[ \mathcal{H}, \mathcal{O}_{\alpha_1} \right] |\psi_{HF}\rangle|$ maximal;

    \item Then the unitary operator is updated to $U_{ex} = e^{\lambda_1^{(1)} \mathcal{O}_{\alpha_1}}$ with $\lambda_1^{(1)}$ determined by minimizing: 
\beq
\langle \psi_{HF}| e^{\lambda_1^{(1)} \mathcal{O}_{\alpha_1}^\dag} \mathcal{H} e^{\lambda_1^{(1)} \mathcal{O}_{\alpha_1}}|\psi_{HF}\rangle;
\eeq

    \item The wave function is updated to $|\psi_{\alpha_1}\rangle = e^{\lambda_1^{(1)} \mathcal{O}_{\alpha_1}}|\psi_{HF}\rangle$, and the expectation energy of the system with respect to the wave function is given by $E_{\alpha_1} = \langle \psi_{\alpha_1} | \mathcal{H} |\psi_{\alpha_1}\rangle$;

    \item We evaluate $\langle \psi_{\alpha_1} | \left[ \mathcal{H}, \mathcal{O}_1 \right] |\psi_{\alpha_1}\rangle$, $\langle \psi_{\alpha_1} | \left[ \mathcal{H}, \mathcal{O}_2 \right] |\psi_{\alpha_1}\rangle$, $\langle \psi_{\alpha_1} | \left[ \mathcal{H}, \mathcal{O}_3 \right] |\psi_{\alpha_1}\rangle$, and $\langle \psi_{\alpha_1} | \left[ \mathcal{H}, \mathcal{O}_4 \right] |\psi_{\alpha_1}\rangle$;

    \item Based on the expectation values evaluated in step 5, we can find the operator $\mathcal{O}_{\alpha_2} \in \mathcal{P}$, which makes $|\langle \psi_{\alpha_1} | \left[ \mathcal{H}, \mathcal{O}_{\alpha_2} \right] |\psi_{\alpha_1}\rangle|$ maximal;

    \item The unitary operator is updated to $U_{ex} = e^{\lambda_2^{(2)} \mathcal{O}_{\alpha_2}} e^{\lambda_1^{(2)} \mathcal{O}_{\alpha_1}}$, with $\lambda_1^{(2)}$ and $\lambda_2^{(2)}$ determined by minimizing:
\beq
\langle \psi_{HF}| e^{\lambda_1^{(2)} \mathcal{O}_{\alpha_1}^\dag} e^{\lambda_2^{(2)} \mathcal{O}_{\alpha_2}^\dag} \mathcal{H} e^{\lambda_2^{(2)} \mathcal{O}_{\alpha_2}} e^{\lambda_1^{(2)}  \mathcal{O}_{\alpha_1}}|\psi_{HF}\rangle;
\eeq

    \item The wave function is updated to $|\psi_{\alpha_2}\rangle = e^{\lambda_2^{(2)} \mathcal{O}_{\alpha_2}} e^{\lambda_1^{(2)}  \mathcal{O}_{\alpha_1}}|\psi_{HF}\rangle$, and the energy expectation of the system with respect to the approximate wave function is $E_{\alpha_2} = \langle \psi_{\alpha_2} | \mathcal{H} |\psi_{\alpha_2}\rangle$.

    \item We calculate $|E_{\alpha_2}-E_{\alpha_1}|$ and, if it is smaller than the convergence criteria $\epsilon$, the exact unitary operator is approximated by:
\beq
U_{ex} = e^{\lambda_2^{(2)} \mathcal{O}_{\alpha_2}} e^{\lambda_1^{(2)}  \mathcal{O}_{\alpha_1}}.
\eeq

    \item Otherwise, we repeat the steps from 5 to 9, until $|E_{\alpha_M}-E_{\alpha_{M-1}}|<\epsilon$. The exact unitary operator is then
\beq
U_{ex} = e^{\lambda_M^{(M)} \mathcal{O}_{\alpha_M}} e^{\lambda_{M-1}^{(M)} \mathcal{O}_{\alpha_{M-1}}} \cdots e^{\lambda_1^{(M)}  \mathcal{O}_{\alpha_1}}.
\eeq

\end{enumerate}

Typically, a small enough $\epsilon$ ensures good convergence to the true ground state. To illustrate this point, we used the quantum simulator provided by qiskit \cite{qiskit2019} to implement the adaptive VQE for the quantum Hall model. In particular, the magnetic BZ is discretized into $3 \times 12$ mesh points, and we performed adaptive VQE for each mometum point. In Fig.~\ref{fig12} we plot the optimized ground state energy versus the true ground state energy by using $\epsilon=0.01$, and the smaller difference between the optimized ground state energy and the true ground state energy ($<0.03$) indicates that good approximations are achieved.

\subsection{Truncating the optimization sequence} \label{appendix_d2}

Now based on the adaptive VQE described in the above, the approximate unitary operator, which can transform the Hartree-Fock ground state to the approximated ground state, is given by:
\beq \label{Eq_full}
U_{ex} = e^{\lambda_M^{(M)} \mathcal{O}_{\alpha_M}} e^{\lambda_{M-1}^{(M)} \mathcal{O}_{\alpha_{M-1}}} \cdots e^{\lambda_1^{(M)}  \mathcal{O}_{\alpha_1}}.
\eeq
However, this state preparation usually is too deep for NISQ machines. Fortunately, the robustness of the Chern number allows us to do a further truncation.

Note that with the full sequence with $U_{ex}$ given by Eq.~(\ref{Eq_full}), the ground state is approximated by:
\beq
|GS^{(M)}\rangle = e^{\lambda_M^{(M)} \mathcal{O}_{\alpha_M}} e^{\lambda_{M-1}^{(M)} \mathcal{O}_{\alpha_{M-1}}} \cdots e^{\lambda_1^{(M)}  \mathcal{O}_{\alpha_1}} |\psi_{HF}\rangle.
\eeq
If we choose a small $\epsilon$ (i.e. $\epsilon=0.01$), we found that $E_{\alpha_M}=\langle GS^{M}| \mathcal{H} |GS^{M}\rangle$ is very close to the true ground state energy $E_{GS}$.

On the other hand, in the process of implementing the adaptive VQE described in the last subsection, at the end of each iteration step we can obtain an expectation energy of the system with respect to the approximate wave function at the given iteration. For example, the expectation energy $E_{\alpha_n}$ corresponds to the wave function
\beq
e^{\lambda_n^{(n)} \mathcal{O}_{\alpha_n}} e^{\lambda_{n-1}^{(n)} \mathcal{O}_{\alpha_{n-1}}} \cdots e^{\lambda_1^{(n)}  \mathcal{O}_{\alpha_1}} |\psi_{HF}\rangle.
\eeq

\begin{figure}[!h]
  \centering
  \includegraphics[width=0.9\columnwidth]{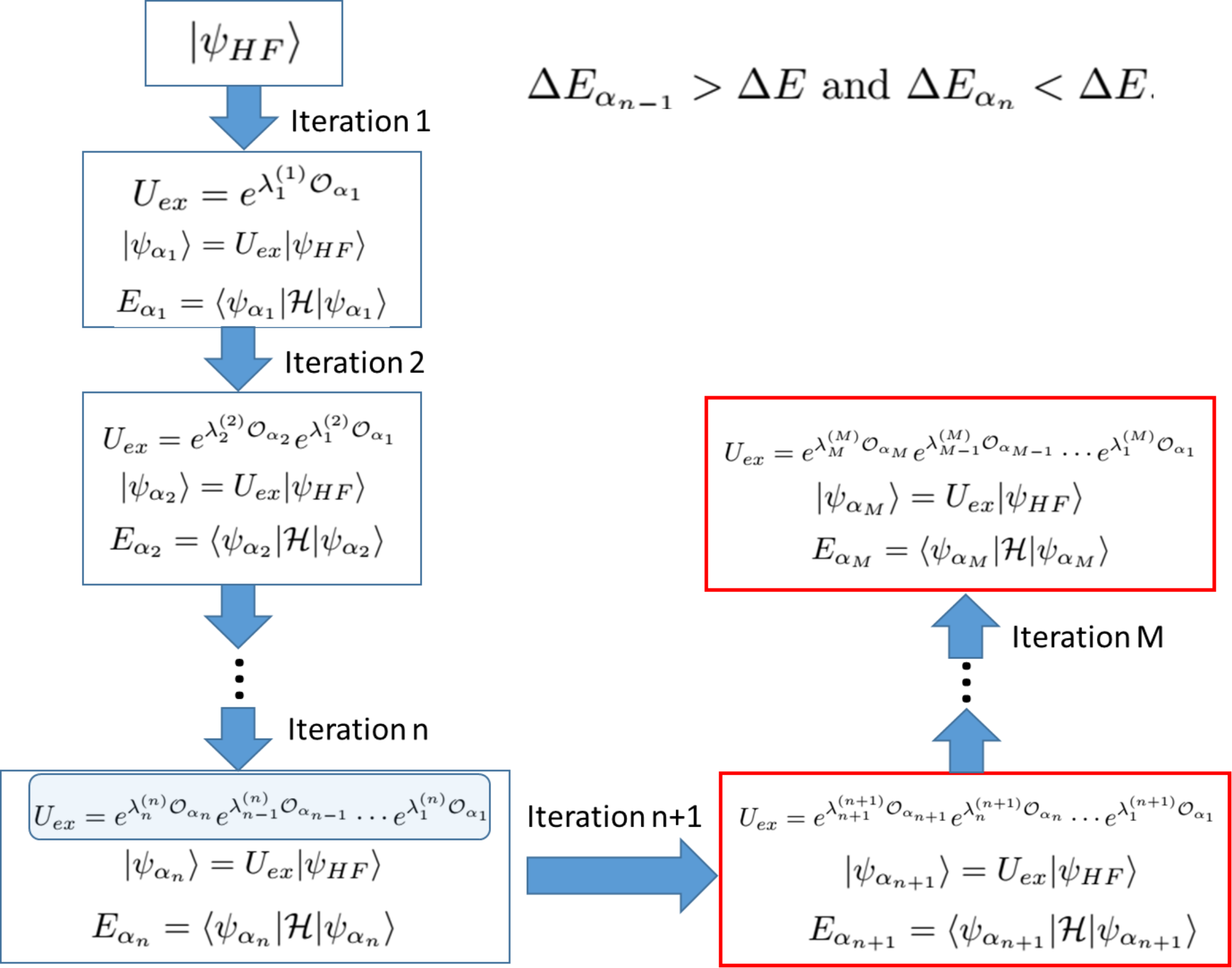}
  \caption{The illustration of the sequence truncation under the condition $\Delta E_{\alpha_{n-1}} > \Delta E$ and $\Delta E_{\alpha_{n}} < \Delta E$. The iterations in red boxes are dropped under the condition.}
   \label{figs4}
\end{figure}

We write these expectation energies obtained in the adaptive VQE as an array: $\left( E_{\alpha_1}, E_{\alpha_2}, \cdots, E_{\alpha_M} \right)$. From the adaptive VQE \cite{Grimsley_NC_2019}, we know that if we can iterate the optimization infinitely, the prepared state should ideally be identical to the true ground state (although this might fail, as the adaptive VQE is not guaranteed to produce the ground state after an infinite number of steps, or it could reach the ground state with a finite number of steps, since the ground state can be formed exactly from a finite number of UCC factors). Moreover, the expectation energy with respect to the wave function obtained after each iteration decreases (hopefully all the way to the true ground state energy). In this sense, how the expectation energy converges to the true ground state energy reflects how the prepared wave function converges to the true ground state wave function. Based on this observation, we can further define:
\begin{align}
\mathbf{\Delta} &= \left( E_{\alpha_1}-E_{\alpha_M}, E_{\alpha_2}-E_{\alpha_M}, \cdots, E_{\alpha_M}-E_{\alpha_M} \right) \nonumber \\
&= \left( \Delta E_{\alpha_1}, \Delta E_{\alpha_2}, \cdots, 0 \right).
\end{align}
Note that $\Delta E_{\alpha_n} \approx E_{\alpha_n} - E_{GS}$, so the $n$th element in $\mathbf{\Delta}$, $\Delta E_{\alpha_n}$, characterizes how far away the wave function obtained after the $n$th iteration is from the true ground state wave function.

Then based on the array $\mathbf{\Delta}$, we can have a controlled way to truncate the full sequence with $U_{ex}$. For example, we can choose a $\Delta E$ so that $\Delta E_{\alpha_{n-1}} > \Delta E$ and $\Delta E_{\alpha_{n}} < \Delta E$. It means that if we use:
\beq
|GS_{trun}\rangle = e^{\lambda_n^{(n)} \mathcal{O}_{\alpha_n}} e^{\lambda_{n-1}^{(n)} \mathcal{O}_{\alpha_{n-1}}} \cdots e^{\lambda_1^{(n)}  \mathcal{O}_{\alpha_1}} |\psi_{HF}\rangle,
\eeq
the prepared ground state energy would be larger than the true ground state energy by a value $\sim\Delta E$. The truncation procedure is illustrated in Fig.~\ref{figs4}. Typically, $\Delta E$ should be much smaller than the exciation gap above the ground state to guarantee the true ground state can be properly prepared. In the calculations shown in the main text, $\Delta E=0.2$ and $\Delta E=0.3$ are much smaller than the exciation gap of the model, which is $\sim2$.

\begin{figure}[!h]
  \centering
  \includegraphics[width=1\columnwidth]{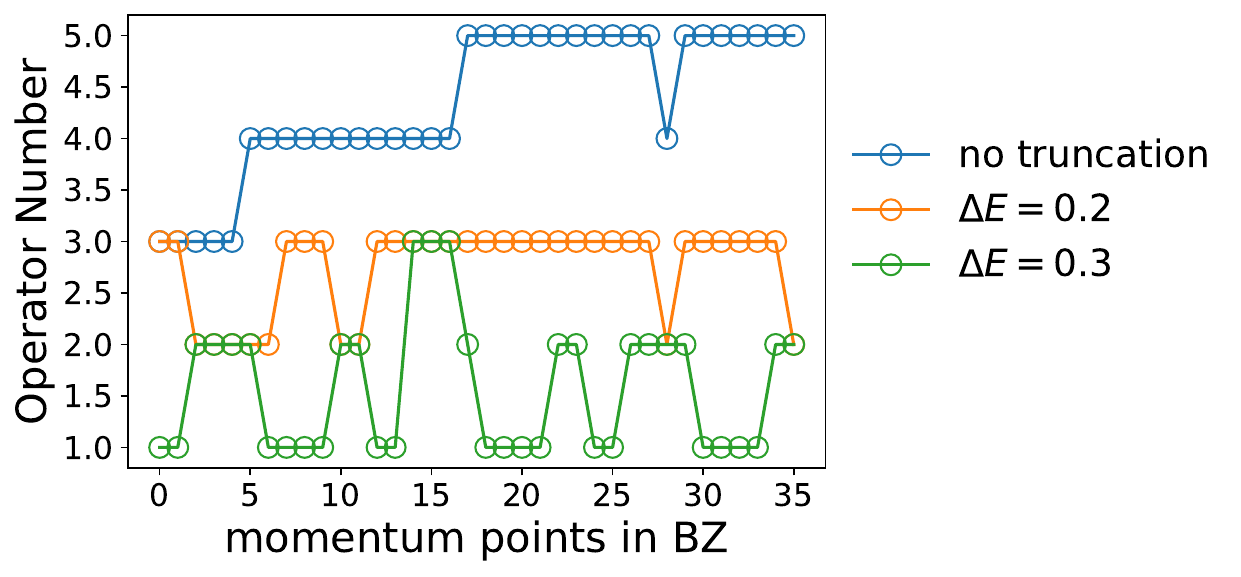}
  \caption{The operator number changes due to the introduction of truncation parameter $\Delta E$. In the calculation the convergence parameter $\epsilon=0.01$, which is the setting in the main text.}
   \label{figs5}
\end{figure}

As we have discussed in the main text, the purpose of the truncation is to reduce the depth of the quantum circuits so that we could run it on the present quantum hardware. The truncation can be done in two ways in the main text: one is to introduce a truncation parameter $\Delta E$; the other is to increase the convergence parameter $\epsilon$. The advantage of introducing a truncation parameter is that the parameter $\Delta E$ is a good measure of how far away the truncated states used for the calculation of topological invariant are from the true ground states. For the flux-$2\pi/3$ quantum hall model, which we have studied in Sec. \ref{sec4a} of the main text, the operation numbers required for each momentum points in the magnetic Brillouin zone for different truncation parameter $\Delta E$ are shown in Fig.~\ref{figs5}.

\section{Projective measurement to replace the Hadamard test} \label{appendix_e}

To have a motivation to introduce projective measurements, let us first rewrite the wave function overlap into the following form:
\beq \label{overlap_proj}
\langle \Psi(\bk) | \Psi(\bk + \delta \bk) \rangle = \langle \Psi(\bk) | \mathfrak{U}(\bk+\delta \bk,\bk) |\Psi(\bk)\rangle,
\eeq
where the operator $\mathfrak{U}(\bk+\delta \bk,\bk)$ transforms the wave function at the momentum point $\bk$ to the wave function at the momentum point $\bk+\delta \bk$, namely:
\beq
\mathfrak{U}(\bk+\delta \bk,\bk) |\Psi(\bk)\rangle = | \Psi(\bk + \delta \bk) \rangle.
\eeq
The new expression of wave function overlaps indicates that except for the calculations of wave function overlaps from the Hadamard test algorithm, we can also calculate them by measuring the expectation values of the operator $\mathfrak{U}(\bk+\delta \bk,\bk)$. The expression of the operator $\mathfrak{U}(\bk+\delta \bk,\bk)$ can be determined by the adaptive VQE algorithm and can be expressed in general in terms of the operators in the operator pools for the corresponding VQE:
\beq \label{evo_op}
\mathfrak{U}(\bk+\delta \bk,\bk) \approx e^{\lambda_N \mathcal{O}_{\alpha_N}} e^{\lambda_{N-1} \mathcal{O}_{\alpha_{N-1}}} \cdots e^{\lambda_1 \mathcal{O}_{\alpha_1}},
\eeq
where $\mathcal{O}_{\alpha_i}$ is an operator in the operator pool of the adaptive VQE, and $\lambda_i$ is the coefficient. The adaptive VQE algorithm is used to determine the orders of the operators in the above expression and also the coefficient.

From Appendix \ref{appendix_d} and \ref{appendix_f}, one can easily identify that the operator $\mathcal{O}_{\alpha_i}$ can be expressed as a linear combination of a series Pauli strings. Explicitly, for a system containing $N$ qubits, the operator can be expressed as $\mathcal{O}_{\alpha_i} = \sum_{j} P_j$, where $P_j = \otimes_n^{N} \sigma_n^{\alpha_n}$ with $\alpha_n=x,y,z,0$. More importantly, one can prove that the  Pauli strings in the sum commute with each other \cite{Mitarai_PRR_2019}. Therefore, we have:
\beq \label{Pauli_str_op}
e^{\lambda_i \mathcal{O}_{\alpha_i}} = \prod_j e^{\lambda_i P_j}  =\prod_j \left( \cos \lambda_i + i \sin \lambda_i P_j \right).
\eeq
Inserting Eq.~(\ref{Pauli_str_op}) into the expression of $\mathfrak{U}(\bk+\delta \bk,\bk)$ given in Eq.~(\ref{evo_op}), one can find the operator $\mathfrak{U}(\bk+\delta \bk,\bk)$ can be written as the linear combination of Pauli strings:
\beq \label{evo_lin}
\mathfrak{U}(\bk+\delta \bk,\bk) = \sum_j^{M} c_j P_j.
\eeq
We can easily find that the number of terms in the summation in Eq.~(\ref{evo_lin}) is bounded by $4^N-1$ with $N$ is the number of qubits ($4$ is due to the fact that for each qubit there are $4$ different Pauli matrices acting on it). Obviously, such a method can not be scaled to large systems, but for small size systems it works well and can be used to reduce circuit depths.

In our calculations for small size systems, we can easily determine the coefficients in front of $P_j$ in Eq.~(\ref{evo_lin}) with a classical algorithm. After correctly measuring the expectation values of $P_j$ by using the wave function at momentum point $\bk$ on quantum computers, we can straightforwardly calculate:
\begin{align}
\langle \Psi(\bk) | \Psi(\bk + \delta \bk) \rangle &= \langle \Psi(\bk) | \mathfrak{U}(\bk+\delta \bk,\bk) |\Psi(\bk)\rangle \nonumber \\
&= \sum_j c_j \langle \Psi(\bk) | P_j |\Psi(\bk)\rangle.
\end{align}

\section{The operator pool for the interacting Chern insulator model} \label{appendix_f}

The operator pool for the interacting Chern insulator model constructed on the two unit cells is still constructed based on the unitary coupled cluster theory, but the caution here is that to account for the effect of interaction, double excitation operators must be included. Based on the Hamiltonian in Eq. (\ref{interacting_Chern_Hamiltonian}), the operators generating single excitations/de-excitations are given by:
\beq
\begin{cases}
\mathcal{O}_1 = c_{\bk;1}^\dag c_{\bk;2} - c_{\bk;2}^\dag c_{\bk;1}, \\
\mathcal{O}_2 = c_{\bk;1}^\dag c_{\bk;3} - c_{\bk;3}^\dag c_{\bk;1}, \\
\mathcal{O}_3 = c_{\bk;1}^\dag c_{\bk;4} - c_{\bk;4}^\dag c_{\bk;1}, \\
\mathcal{O}_4 = c_{\bk;2}^\dag c_{\bk;3} - c_{\bk;3}^\dag c_{\bk;2}, \\
\mathcal{O}_5 = c_{\bk;2}^\dag c_{\bk;4} - c_{\bk;4}^\dag c_{\bk;2}, \\
\mathcal{O}_6 = c_{\bk;3}^\dag c_{\bk;4} - c_{\bk;4}^\dag c_{\bk;3}, \\
\mathcal{O}_7 = i(c_{\bk;1}^\dag c_{\bk;2} + c_{\bk;2}^\dag c_{\bk;1}), \\
\mathcal{O}_8 = i(c_{\bk;1}^\dag c_{\bk;3} + c_{\bk;3}^\dag c_{\bk;1}), \\
\mathcal{O}_9 = i(c_{\bk;1}^\dag c_{\bk;4} + c_{\bk;4}^\dag c_{\bk;1}), \\
\mathcal{O}_{10} = i(c_{\bk;2}^\dag c_{\bk;3} + c_{\bk;3}^\dag c_{\bk;2}), \\
\mathcal{O}_{11} = i(c_{\bk;2}^\dag c_{\bk;4} + c_{\bk;4}^\dag c_{\bk;2}), \\
\mathcal{O}_{12} = i(c_{\bk;3}^\dag c_{\bk;4} + c_{\bk;4}^\dag c_{\bk;3}),
\end{cases}
\eeq
where the first $6$ terms are the operators obtained by the conventional unitary coupled cluster theory, and the last $6$ terms are introduced to take into account the spin-orbit coupling terms. These terms can be written as Pauli strings similar to Eq. (\ref{JW}) by using the Jordan-Wigner transformation, and we do not write them down explicitly here.

The operators generating double excitations/de-excitations are obtained directly by following the unitary coupled cluster theory:
\beq
\begin{cases}
\mathcal{O}_{13} = (c_{\bk;1}^\dag c_{\bk;2} c_{\bk;3}^\dag c_{\bk;4} - c_{\bk;4}^\dag c_{\bk;3} c_{\bk;2}^\dag c_{\bk;1}), \\
\mathcal{O}_{14} = (c_{\bk;1}^\dag c_{\bk;3} c_{\bk;2}^\dag c_{\bk;4} - c_{\bk;4}^\dag c_{\bk;2} c_{\bk;3}^\dag c_{\bk;1}), \\
\mathcal{O}_{15} = (c_{\bk;1}^\dag c_{\bk;4} c_{\bk;2}^\dag c_{\bk;3} - c_{\bk;3}^\dag c_{\bk;2} c_{\bk;4}^\dag c_{\bk;1}), \\
\mathcal{O}_{16} = (c_{\bk;1}^\dag c_{\bk;2} c_{\bk;4}^\dag c_{\bk;3} - c_{\bk;3}^\dag c_{\bk;4} c_{\bk;2}^\dag c_{\bk;1}), \\
\mathcal{O}_{17} = (c_{\bk;1}^\dag c_{\bk;3} c_{\bk;4}^\dag c_{\bk;2} - c_{\bk;2}^\dag c_{\bk;4} c_{\bk;3}^\dag c_{\bk;1}), \\
\mathcal{O}_{18} = (c_{\bk;1}^\dag c_{\bk;4} c_{\bk;3}^\dag c_{\bk;2} - c_{\bk;2}^\dag c_{\bk;3} c_{\bk;4}^\dag c_{\bk;1}). \\
\end{cases}
\eeq
To write these operators in terms of Pauli strings, we need to use the fact that:
\begin{align}
c_n^\dag c_{n+m} &= \frac{1}{4} \left( XX_{n;n+m} + YY_{n;n+m} \right) \nonumber \\
&+ \frac{i}{4} \left( XY_{n;n+m} - YX_{n;n+m} \right),
\end{align}
where the subscripts are ordered along the Jordan-Wigner string, $XX_{n;n+m}$, $YY_{n;n+m}$, $XY_{n;n+m}$ and $YX_{n;n+m}$ are Pauli strings and have the following explicit expressions:
\beq
\begin{cases}
XX_{n;n+m} = \sigma_n^x \prod_{n<j<n+m} \sigma_j^z \sigma_{n+m}^x, \\ 
YY_{n;n+m} = \sigma_n^y \prod_{n<j<n+m} \sigma_j^z \sigma_{n+m}^y, \\
XY_{n;n+m} = \sigma_n^x \prod_{n<j<n+m} \sigma_j^z \sigma_{n+m}^y, \\
YX_{n;n+m} = \sigma_n^y \prod_{n<j<n+m} \sigma_j^z \sigma_{n+m}^x.
\end{cases}
\eeq
Therefore, we have generically:
\begin{align} \label{Eq:double_excitation}
c_{n}^\dag c_m c_i^\dag c_j - c_j^\dag c_i c_{m}^\dag c_n &= \frac{i}{8} \left( XX_{n;m} XY_{i;j} - XX_{n;m} YX_{i;j} \right) \nonumber \\
&+ \frac{i}{8} \left( YY_{n;m} XY_{i;j} - YY_{n;m} YX_{i;j} \right) \nonumber \\
&+ \frac{i}{8} \left( XY_{n;m} XX_{i;j} + XY_{n;m} YY_{i;j} \right) \nonumber \\
&- \frac{i}{8} \left( YX_{n;m} XX_{i;j} + YX_{n;m} YY_{i;j} \right).
\end{align}
For the interacting Chern insulator model considered in the main text, the quantum circuits representing the $8$ terms in Eq.~(\ref{Eq:double_excitation}) have the similar structures, details of which were provided in \cite{Fedorov_arXiv_2021}. The circuit depth for each of them is $5$, so the depth for an operator generating double excitations/de-excitations will be $40$.

\section{Calibration data and other measurements from quantum hardware} \label{appendix_g}

\subsection{Calibration data of IBMQ-Toronto}

The measured Chern numbers shown in Fig.~\ref{fig2} were obtained on IBMQ-Toronto. The quantum calculations were repeated two times on 08/03/2020 and 10/16/2020. The calibration data for the relevant qubits where the calculations were carried out are summarized in Table \ref{tab:table-calibration_toronto1} and Table \ref{tab:table-calibration_toronto2}.

\begin{table}[h]
 \resizebox{0.48\textwidth}{!}{\begin{tabular}{|| c | c | c | c | c || c | c ||} 
 \hline
 \backslashbox{qubit}{gate} & Id & $U_1$ & $U_2$ & $U_3$ & \backslashbox{qubit}{gate} & CX  \\ [0.2ex] 
 \hline\hline
 1 & $1.64\times10^{-4}$ & 0 & $1.64\times10^{-4}$ & $3.27\times10^{-4}$ & [1,4] & $9.36\times10^{-3}$ \\ 
 \hline
 4 & $3.07\times10^{-4}$ & 0 & $3.07\times10^{-4}$ & $6.14 \times10^{-4}$ & [4,7] & $1.16 \times 10^{-2}$ \\ 
 \hline
 7 & $4.49\times10^{-4}$ & 0 & $3.07\times10^{-4}$ & $8.98 \times10^{-4}$ & & \\ [0.5ex] 
 \hline
\end{tabular}}
\caption{\label{tab:table-calibration_toronto1} The calibration error data on 08/03/2020, for the gates operated on qubits $1$, $4$  and $7$ on  IBMQ-Toronto.}
\end{table}

\begin{table}[!h]
 \resizebox{0.48\textwidth}{!}{\begin{tabular}{||c | c | c | c | c || c | c ||} 
 \hline
 \backslashbox{qubit}{gate} & Id & $U_1$ & $U_2$ & $U_3$ & \backslashbox{qubit}{gate} & CX  \\ [0.5ex] 
 \hline\hline
 0 & $3.18\times10^{-4}$ & 0 & $3.18\times10^{-4}$ & $6.35\times10^{-4}$ & [0,1] & $6.71\times10^{-3}$ \\ 
 \hline
 1 & $1.86\times10^{-4}$ & 0 & $1.86\times10^{-4}$ & $3.72 \times10^{-4}$ & [1,4] & $9.13 \times 10^{-3}$ \\ 
 \hline
 4 & $2.06\times10^{-4}$ & 0 & $2.06\times10^{-4}$ & $4.13 \times10^{-4}$ & & \\ [1ex] 
 \hline
\end{tabular}}
\caption{\label{tab:table-calibration_toronto2} The calibration error data on 10/16/2020, for the gates operated on qubits $0$, $1$  and $4$ on  IBMQ-Toronto.}
\end{table}

\subsection{Calibration data on IBMQ-Montreal and other measurements on this machine} 

The measured Chern number shown in Fig.~\ref{fig6} of the main text was obtained on IBMQ-Montreal. The calibration data for the machine, when the calculations were carried out, is summarized in Table \ref{tab:table-calibration_montreal} below for the qubits on which the calculations were performed.

\begin{table}[!h]
  \resizebox{0.48\textwidth}{!}{\begin{tabular}{||c | c | c | c | c || c | c ||} 
 \hline
 \backslashbox{qubit}{gate} & Id & RZ & SX & X & \backslashbox{qubit}{gate} & CX  \\ [0.5ex] 
 \hline\hline
 8 & $2.23\times10^{-4}$ & 0 & $2.23\times10^{-4}$ & $2.23\times10^{-4}$ & [8,11] & $7.19\times10^{-3}$ \\ 
 \hline
 11 & $1.98\times10^{-4}$ & 0 & $1.98\times10^{-4}$ & $1.98\times10^{-4}$ & [11,14] & $6.33 \times 10^{-3}$ \\ 
 \hline
 14 & $2.90\times10^{-4}$ & 0 & $2.90\times10^{-4}$ & $2.90\times10^{-4}$ & & \\ [1ex] 
 \hline
\end{tabular}}
\caption{\label{tab:table-calibration_montreal} The calibration error data on 05/28/2021, for the gates operated on qubits $8$, $11$  and $14$ on IBMQ-Montreal.}
\end{table}

The measurements of the Chern number of the two-body ground state of the quantum Hall model was repeated $5$ times for $\Delta E=0.2$ and $\Delta E=0.3$ on IBMQ-Montreal, respectively. For the $\Delta E=0.2$ cases, $4$ out of $5$ measurements have the same distribution of the integer-valued field $n(\bk)$ in the magnetic BZ as that shown in Fig.~\ref{fig3}(d), and the remaining one measurement has the distribution of the integer-valued field $n(\bk)$ the same with that shown in Fig.~\ref{fig3}(e). For the $\Delta=0.3$ case, $4$ out of $5$ measurements have the same distribution of the integer-valued field $n(\bk)$ in the magnetic BZ as that shown in Fig.~\ref{fig3}(e), and the remaining one measurement shows the same distribution of the integer-valued field $n(\bk)$ with that shown in Fig.~\ref{fig3}(d).

\subsection{Calibration data on IBMQ-Montreal for the calculation of interacting models}

The measured Chern number shown in Fig.~\ref{fig7} of the main text was obtained on IBMQ-Montreal, and the calculation were repeated twice. The calibration data for the machine, when the calculations were carried out, is summarized in Table \ref{tab:table-calibration_montreal_interacting} below for the qubits on which the calculations were performed.

\begin{table}[!h]
  \resizebox{0.48\textwidth}{!}{\begin{tabular}{||c | c | c | c | c || c | c ||} 
 \hline
 \backslashbox{qubit}{gate} & Id & RZ & SX & X & \backslashbox{qubit}{gate} & CX  \\ [0.5ex] 
 \hline\hline
 0 & $5.53\times10^{-4}$ & 0 & $5.53\times10^{-4}$ & $5.53\times10^{-4}$ & [0,1] & $1.83\times10^{-2}$ \\ 
 \hline
 1 & $7.43\times10^{-4}$ & 0 & $7.43\times10^{-4}$ & $7.43\times10^{-4}$ & [1,4] & $1.86 \times 10^{-2}$ \\ 
 \hline
 4 & $3.97\times10^{-4}$ & 0 & $3.97\times10^{-4}$ & $3.97\times10^{-4}$ & & \\ [1ex] 
 \hline
\end{tabular}}
\caption{\label{tab:table-calibration_montreal_interacting} The calibration error data on 07/06/2021, for the gates operated on qubits $0$, $1$  and $4$ on IBMQ-Montreal.}
\end{table}

\end{document}